\documentclass[aps,prc,preprint,groupedaddress,nofootinbib]{revtex4-2}

\usepackage[utf8]{inputenc}
\usepackage[T1]{fontenc}
\usepackage{lmodern}
\usepackage{amsmath, amssymb, amsfonts}
\usepackage{graphicx}
\usepackage{dcolumn}  
\usepackage{bm}       
\usepackage{hyperref} 
\usepackage{physics}  
\usepackage{siunitx}  
\usepackage{xcolor}   
\usepackage{comment}  

\usepackage[]{mathrsfs}
\usepackage{float}
\usepackage{array}

\begin{document}

\title{Relativistic Second Random-Phase Approximation: Formalism and Applications to Giant Resonances}

\author{D. Vale}
\email{dvale@iv.hr}
\affiliation{Istrian University of Applied Sciences, Preradovićeva 9D, 52100 Pula, Croatia}

\author{G. Saxena}
\affiliation{Department of Physics (H \& S), Govt. Women Engineering College, Ajmer-305002, India}

\author{N. Paar}
\email{npaar@phy.hr}
 \affiliation{Department of Physics, Faculty of Science, University of Zagreb, Bijeni\v cka cesta 32, 10000 Zagreb, Croatia}

\date{\today}

\begin{abstract}
We develop a self-consistent relativistic second random-phase approximation (RSRPA) based on covariant energy density functional with density dependent point-coupling interaction. The formalism extends the conventional simple configuration space composed of particle-hole (\(p\text{–}h\)) transitions by including two-particle–two-hole (\(2p\text{–}2h\)) configurations through a generalized Thouless ansatz and a time-dependent variational principle. To ensure consistency with the underlying mean-field ground state and avoid the double counting of correlations, we implement the subtraction method, in which the static contribution is removed from the residual interaction.  The first applications of the RSRPA include studies of isoscalar and isovector giant monopole (ISGMR and IVGMR) and quadrupole resonances (ISGQR and IVGQR) in $^{16}$O, which demonstrate the importance of \(2p\text{–}2h\) couplings to describe the fragmentation of the transitions strength and the stabilizing role of subtraction method.
\end{abstract}

\maketitle

\section{Introduction}

The studies of nuclear excitations and collective modes within self-consistent frameworks have undergone significant development over the past decades \cite{ Paar2003, Paar2004, Litvinova2007_PLB, Dao, Litvinova2008, Litvinova2008b, Colo2010,MishevVoronov2010, Litvinova2011, Niu2012, Reinhard2014,  Litvinova2023, Knapp, Vesely:2025, bender2003self, Garg2018, paar2007exotic, Savran2013}. Both the non-relativistic and relativistic random-phase approximation (RPA and RRPA, respectively), built upon the density functional theory (DFT), have proven successful in describing the properties of giant resonances, such as centroid energies, transition strengths, and isospin dependencies across the nuclear chart. However, standard (R)RPA is limited to configurations built from single particle–hole (\(1p\text{–}1h\)) excitations, and therefore cannot account for effects arising from more complex correlations. In particular, it fails to describe the spreading width (\(\Gamma^\downarrow\)) of giant resonances and the fragmentation of strength associated with coupling to two-particle–two-hole (\(2p\text{–}2h\)) states \cite{Baldwin1947, Harakeh1976, Youngblood1977, Ulrikson1977, Youngblood1997, Lui2001, Youngblood2001, Hartmann2002, Tamii2011, Savran2013, HOWARD2020}.
This limitation motivates the development of extended approaches that include higher-order configurations beyond the standard RPA.

The second random-phase approximation (SRPA) provides a natural extension by incorporating both \(1p\text{–}1h\) and \(2p\text{–}2h\) configurations into the excitation operator \cite{Providencia}. This allows for a more realistic description of nuclear dynamics, including configuration mixing, fragmentation, and dynamical self-energy corrections. In non-relativistic implementations, SRPA has been employed with various effective interactions such as Skyrme \cite{Gam1, Gambacurta2011, Minato2016, Yang2021} and Gogny functionals \cite{Gambacurta2012}, showing improvements in the description of strength distributions and low-lying states.
In addition to the approaches discussed above, the SRPA formalism has also been employed in combination with realistic nucleon-nucleon (N–N) interactions. In particular, a renormalized form of the Argonne V18 potential has been used, where short-range correlations are incorporated via the Unitary Correlation Operator Method \cite{Papakonstantinou2007, Papakonstantinou2010}. However, these models often encounter issues with stability, double counting of correlations, and infrared divergences. To address these issues, the subtraction method has been proposed, in which the static limit of the energy-dependent self-energy is removed to ensure consistency with the mean-field ground state \cite{Tselyaev2013, Gambacurta2015}.

A variety of advanced theoretical frameworks have been proposed to account for additional physical effects in nuclear dynamics. The finite Fermi system theory, for instance, accommodates both the single-particle continuum and the coupling to low-lying collective modes~\cite{Caurier2005, Donno2011}. Low-energy nuclear structure can be further refined through particle–vibration and particle–phonon coupling schemes, which describe the interaction between single-particle states and collective phonon excitations~\cite{Colo2010, MishevVoronov2010, Niu2012, Niu2015}. Microscopic multiphonon methods, including the quasiparticle phonon model (QPM), explicitly treat couplings among one-particle–one-hole, two-particle–two-hole, and three-particle–three-hole configurations, offering a means to explore the impact of higher-order correlations on the structure and fragmentation of nuclear collective modes~\cite{Soloviev1992, Ponomarev1996, Ponomarev2001, Shevchenko2004, DeGregorio2016, DeGregorio2018, DeGregorio2021}. In addition, extended random phase approximation (ERPA) approaches have been formulated within the time-dependent density matrix framework to go beyond simple particle–hole excitations. These models describe excited states on a correlated ground state in a symmetry-preserving way and maintain the energy-weighted sum rules~\cite{Tohyama2004,TohyamaSchuck2004,TohyamaTakaharaSchuck2004_EPJA,Tohyama2007,TohyamaUmar2016_PRC,dickhoff2004self}.

Within a relativistic setting, the relativistic quasiparticle time-blocking approximation (RQTBA) incorporates particle–vibration coupling together with two-quasiparticle--phonon configurations, enabling the treatment of giant-resonance fragmentation and damping, while the time-blocking procedure ensures that double counting is avoided~\cite{Litvinova2008, Litvinova2008b, Litvinova2023}. Extensions of the method include multiphonon RQTBA~\cite{Litvinova2015multiphonon} and finite-temperature generalisations~\cite{Litvinova2018FT}, which enlarge the configuration space to account for richer 2p–2h dynamics and broaden the applicability to hot nuclei and astrophysical environments.

While SRPA has been extensively studied in the non-relativistic regime, its relativistic counterpart remains much less explored. Initial investigations within the relativistic framework have been limited to the Tamm–Dancoff approximation, specifically in its diagonal form, which neglects the interaction between two-particle–two-hole ($2p$--$2h$) pairs \cite{ValePaar2024}. On the other hand, a major challenge is to account for the Dirac structure of single-particle states and the residual interaction in the covariant formalism, while ensuring that calculations remain tractable in large $2p$--$2h$ configuration spaces. A covariant SRPA framework is crucial for studying isoscalar and isovector giant resonances, spin-isospin responses, and damping mechanisms in nuclei across the chart. Moreover, such a framework would allow a unified treatment of complex excitations in both ground and excited states, in line with modern goals of predictive nuclear structure theory. Therefore,
in this work, we develop a self-consistent relativistic SRPA (RSRPA) based on the relativistic nuclear energy density functional formulated using point-coupling interaction between nucleons, and apply it in the first calculations of monopole and quadrupole transition strength distributions in the case of $^{16}$O.
This allows us to explore the role of $2p$--$2h$ correlations and their impact on the fragmentation and spreading widths of giant resonances within a covariant theoretical framework and to test the predictive power of this approach against available experimental data \cite{Harakeh1976, Lui2001}. 

The paper is organized as follows.
Section~\ref{sec:thouless} introduces the relativistic form of Thouless theorem and its extension to \(2p\text{--}2h\) space. In Section~\ref{sec:variational}, we apply a time-dependent variational principle to derive the linearized equations of motion. The relativistic extension of the density matrix up to the second order is presented in Section~\ref{sec:relden}. The matrix structure of the RSRPA eigenvalue problem and explicit forms of the residual interaction are discussed in Sections~\ref{sec:matrix_structure} and~\ref{sec:interaction}, respectively. Section~\ref{sec:transitions} presents the calculation of transition strengths and the normalization of excited states, followed by the description of the subtraction method in the Section \ref{sec:subtraction}. Numerical calculations of the response function for the isoscalar and isovector giant monopole (ISGMR and IVGMR) and quadrupole resonance (ISGQR and IVGQR) are shown in Section~\ref{sec:results}. Finally, conclusions and perspectives for further developments are given in Section~\ref{sec:conclusions}.

\section{Formalism}\label{formalism}

\subsection{The extension of Thouless theorem}\label{sec:thouless}
As a starting point for the relativistic linear response formalism in RPA, one considers the fact that an excited Slater determinant can be expressed as the exponential of a particle–hole ($p\text{--}h$) and antiparticle-hole ($\alpha\text{--}h$) excitation operator acting on a reference Hartree (Hartree-Fock) state (for details about the necessity of the inclusion of $\alpha\text{--}h$ pairs in relativistic models see Ref.~\cite{Niksic2002}). This idea stems from the relativistic extension of the original Thouless theorem \cite{Thouless1961}, which remains formally valid under the assumption that the reference state \(|\Phi_0\rangle\) is a Slater determinant built from positive-energy solutions of the Dirac equation.
The time-dependent many-body wave function describing small-amplitude collective excitations around the ground state is then approximated by
\begin{equation}
|\Phi(t)\rangle = \exp\left( \sum_{ph} Z_{ph}(t)\, a_p^\dagger a_h + \sum_{\alpha h} Z_{\alpha h}(t)\, a_\alpha^\dagger a_h \right) |\Phi_0\rangle,
\label{eq:thouless_ansatz}
\end{equation}
where \(a_p^\dagger a_h\) promotes a nucleon from an occupied (hole) state \(h\) to an unoccupied (particle) state \(p\), while $a_\alpha^\dagger a_h$ promotes a nucleon from occupied (hole) state to an antiparticle state in the Dirac sea. The complex amplitudes \(Z_{ph}(t)\) and  \(Z_{\alpha h}(t)\)  encode the time evolution of the excitation and are assumed to be small in the linear response regime.
This parameterization allows for a variational treatment of the system dynamics within the subspace of Slater determinants connected to \(|\Phi_0\rangle\) through single \(p\text{--}h\) or \(\alpha\text{--}h\) excitations. The use of Eq.~\eqref{eq:thouless_ansatz} thus forms the basis for the time-dependent relativistic Hartree (RH) or Hartree-Fock (RHF) approach and, upon linearization, leads directly to the relativistic random-phase approximation (RRPA). It is important to emphasize that this formal structure carries over from the non-relativistic case, provided the creation and annihilation operators act on a Fock space built from Dirac spinors and satisfy the standard fermionic anticommutation relations \cite{mattuck1976guide, greiner2000relativistic}.

To describe the coupling between single-particle and more complex excitations beyond the $1p\text{--}1h$ ($1\alpha\text{--}1h$) pairs, it is necessary to extend the Thouless operator to include higher-order correlations. Specifically, the excited Slater determinant must account for both \(1p\text{--}1h\) (\(1\alpha\text{--}1h\)) and \(2p\text{--}2h\) configurations. This leads to a generalized excitation operator of the form
\begin{equation}
\hat{Z}(t) = \sum_{ph} Z^{(1)}_{ph}(t)\, a_p^\dagger a_h + \sum_{\alpha h} Z_{\alpha h}^{(1)}(t)\, a_\alpha^\dagger a_h
+ \frac{1}{4} \sum_{pp'hh'} Z^{(2)}_{pp'hh'}(t)\, a_p^\dagger a_{p'}^\dagger a_{h'} a_h ,
\label{eq:thouless_extended}
\end{equation}
where \(Z^{(1)}_{ph}(t)\), \(Z^{(1)}_{\alpha h}(t)\) and \(Z^{(2)}_{pp'hh'}(t)\) are the time-dependent amplitudes corresponding to \(1p\text{--}1h\), \(1\alpha\text{--}1h\)  and \(2p\text{--}2h\) excitations, respectively. The factor \(1/4\) in the third term compensates for the double counting of antisymmetric $2p\text{--}2h$ terms. In other words, the amplitudes \(Z^{(2)}_{pp'hh'}\) are defined to be antisymmetric under exchange of particle or hole indices, i.e., \(Z^{(2)}_{pp'hh'} = -Z^{(2)}_{p'p hh'} = -Z^{(2)}_{pp' h' h}\), ensuring uniqueness and correct normalization of the configuration space \cite{Providencia}. The full wave function of a system in the arbitrary state is then expressed as
\begin{equation}
|\Phi(t)\rangle = \exp\left( \hat{Z}(t) \right) |\Phi_0\rangle.
\label{eq:full_ansatz}
\end{equation}
This generalized ansatz spans a variational space that includes not only Slater determinants differing by a single particle–hole (antiparticle-hole) pair but also those obtained by two-particle–two-hole excitations. Such an extension is essential for the consistent derivation of the second RPA (SRPA) formalism, both in the non-relativistic and relativistic context, particularly when aiming to describe the fragmentation of collective modes and the coupling to higher configurations ($2p\text{--}2h$, $3p\text{--}3h$). 

\subsection{Time-dependent variational principle}\label{sec:variational}

To derive the time evolution of the excitation amplitudes \(Z^{(1)}_{ph}(t)\), \(Z^{(1)}_{\alpha h}(t)\), and \(Z^{(2)}_{pp'hh'}(t)\), we employ the time-dependent variational principle. In particular, we adopt the symmetrized form of the Dirac–Frenkel variational approach (see Refs. \cite{DeAngelis1991, Raab2000}), requiring that the following action functional is stationary:
\begin{equation}
\mathcal{S} [Z^{(1)}, Z^{(2)}, Z^{(1)*}, Z^{(2)*}]=\frac{1}{2} \int d t\left(\langle\Phi(t)| i \partial_t-\hat{H}|\Phi(t)\rangle+\langle\Phi(t)|-i \overleftarrow{\partial_t}-\hat{H}|\Phi(t)\rangle\right)
\label{eq:variational_principle}
\end{equation}
The stationarity condition is imposed on the action functional \(\mathcal{S}[Z^{(1)}, Z^{(2)}, Z^{(1)*}, Z^{(2)*}]\), defined in Eq.~\eqref{eq:variational_principle}, by performing independent variations with respect to the excitation amplitudes \(Z^{(1)}\), \(Z^{(2)}\), and their complex conjugates \(Z^{(1)*}\), \(Z^{(2)*}\). The trial wave function is defined by Eq.~\eqref{eq:full_ansatz}, with the excitation operator \(\hat{Z}(t)\) given in Eq.~\eqref{eq:thouless_extended}, incorporating  \(1p\text{--}1h\), \(1\alpha\text{--}1h\) and \(2p\text{--}2h\) excitations. The Hamiltonian density is systematically derived by performing a Legendre transformation of the Lagrangian density associated with the quantum mechanical system. The Lagrangian density which involves point-coupling interaction is described in the Section \ref{sec:interaction}.
In the limit of small-amplitude motion, we expand the exponential operator up to the second order:
\begin{equation} 
|\Phi(t)\rangle=\left(1+\hat{Z}(t)+\frac{1}{2} \hat{Z}^2(t)\right)\left|\Phi_0\right\rangle+\mathcal{O}\left(\hat{Z}^3\right),
\label{expanded2}
\end{equation}
Configurations of the $1p\text{--}1\alpha\text{--}2h$ and $2\alpha\text{--}2h$ type are not considered in this work, in order to have the dimension of the  RSRPA matrix size feasible for calculations. 

Separating static and dynamic parts of amplitudes  \(Z^{(1)}_{ph}(t)\), \(Z^{(1)}_{\alpha h}(t)\) and \(Z^{(2)}_{pp'hh'}(t)\) similar to \cite{Providencia} leads to:
\begin{equation}
\begin{aligned}
Z^{(1)}_{ph}(t) =& C^{(1)}_{ph} + {\chi}^{(1)}_{ph}(t,)\\
Z^{(1)}_{\alpha h}(t) =& C^{(1)}_{\alpha h} + \chi^{(1)}_{\alpha h}(t),\\
Z^{(2)}_{pp'hh'}(t) =& C^{(2)}_{pp'hh'}+ \chi^{(2)}_{pp'hh'}(t),\quad\text{and}\quad c. c.
\end{aligned}    
\end{equation}
In the Hartree or Hartree--Fock approximation, the static components of the amplitudes, denoted by \( C^{(1)}_{ph} \) and \( C^{(1)}_{\alpha h} \), vanish as a consequence of Brillouin's theorem, which states that the Hamiltonian does not couple the uncorrelated ground state to particle--hole excitations \cite{BlaizotRipka1986, Suhonen2007}. Correlations involving two-particle--two-hole (2p--2h) configurations are in most cases not included within the Hartree or Hartree--Fock framework, and the corresponding coefficients \( C^{(2)}_{pp'hh'} \) are therefore neglected here. As a result, the nuclear ground state remains unaffected by such configurations at the mean-field level, and only the dynamical parts \( \chi(t) \) contribute to the response \cite{Providencia}.
Further substituting Eq.~\eqref{expanded2} into Eq.~\eqref{eq:variational_principle}, while imposing stationarity with respect to all independent variations, yields a set of coupled linearized first order differential equations for \(\chi^{(1)}(t)\) and \(\chi^{(2)}(t)\): 
\begin{equation}
\begin{aligned}
i\hbar\dot{\chi}_1(t)~&=~A_{1,1} \chi_1(t) + ... + A_{1, n} \chi_n(t) + B_{1,1} \chi_1^{*}(t) + ... + B_{1,n} \chi_n^{*}(t)\\
i\hbar\dot{\chi}_2(t)~&=~A_{2,1} \chi_1(t) + ... + A_{2, n} \chi_n(t) + B_{2,1} \chi^{*}_1(t) + ... + B_{2,n} \chi_n^{*}(t)
\\
\vdots\\
i\hbar\dot{\chi}_n(t)~&=~A_{n,1} \chi_1(t) + ... + A_{n, n} \chi_n(t) + B_{n,1} \chi_1^{*}(t) + ... + B_{n,n} \chi_n^{*}(t)
\\
-i\hbar\dot{\chi}_1^{*}(t)~&=~B_{1,1}^{*} \chi_1(t) + ... + B_{1, n}^{*} \chi_n(t) + A_{1,1}^{*} \chi_1^{*}(t) + ... + A_{1,n}^{*} \chi_n^{*}(t)
\\
-i\hbar\dot{\chi}_2^{*}(t)~&=~B_{2,1}^{*} \chi_1(t) + ... + B_{2, n}^{*} \chi_n(t) + A_{2,1}^{*} \chi_1^{*}(t) + ... + A_{2,n}^{*} \chi_n^{*}(t)
\\
\vdots\\
-i\hbar\dot{\chi}_n^{*}(t)~&=~B_{n,1}^{*} \chi_1(t) + ... + B_{n, n}^{*} \chi_n(t) + A_{n,1}^{*} \chi_1^{*}(t) + ... + A_{n,n}^{*} \chi_n^{*}(t),
\end{aligned}  
\label{coupled}
\end{equation}
where we used a compact notation in which the index \( n \) runs from \( 1 \) to \( N_{ph} \) for $1p$--$1h$ amplitudes, from \( N_{ph}+1 \) to \( N_{ph} + N_{\alpha h} \) for 1\(\alpha\)--$1h$ amplitudes, and from \( N_{ph} + N_{\alpha h} + 1 \) to \( N_{\text{tot}} \) for $2p$--$2h$ amplitudes. Here, \( N_{ph} \) denotes the total number of $1p$--$1h$ configurations, \( N_{\alpha h} \) the number of 1\(\alpha\)--$1h$ configurations, and \( N_{2p2h} \) the number of $2p$--$2h$ configurations, with
\begin{equation}
N_{ph} + N_{\alpha h} + N_{2p2h} = N_{\text{tot}}.
\end{equation}
Assuming harmonic time dependence of dynamic parts, i.e., 
\begin{equation}
\begin{aligned}
\chi^{(1)}_{ph}(t) = X^{(1)}_{ph} e^{-i\omega t}& + Y^{(1)*}_{ph} e^{+i\omega t}, \quad \chi^{(1)}_{\alpha h}(t) = X^{(1)}_{\alpha h} e^{-i\omega t} + Y^{(1)*}_{\alpha h} e^{+i\omega t}, \\
\chi^{(2)}_{pp'hh'}(t) =& X^{(2)}_{pp'hh'} e^{-i\omega t} + Y^{(2)*}_{pp'hh'} e^{+i\omega t}, \quad c.c.
\end{aligned}
\end{equation}
Eq. \eqref{coupled} may be written in the following form:
\begin{align}
\hbar \omega & X_n^\omega \exp\left( -i \omega t \right) - \hbar \omega Y_n^{\omega{} *} \exp\left( i \omega t\right) ~=~ \sum_k A_{n k} \left\lbrace X_k^\omega \exp\left( - i\omega t\right)\right.  \nonumber\\ ~&+~ \left. Y_k^{\omega {}*} \exp\left( i\omega t\right) \right\rbrace 
+~ \sum_{k}B_{n k}\left\lbrace X_k^{\omega{}*} \exp\left( i\omega t \right)~+~ Y_k^\omega \exp\left( -i\omega t \right)\right\rbrace
\end{align}
while keeping the same notation for indices $n$ and $k$ for both matrices $A$ and $B$ and amplitudes $X$ and $Y$ as in Eq. \eqref{coupled}. For the complex conjugate solutions we have:
\begin{align}
\hbar \omega & X_n^{\omega *} \exp\left( i \omega t \right) -\hbar \omega Y_n^\omega \exp\left( -i \omega t\right) ~=~ \sum_k B_{n k}^{*} \left\lbrace X_k^\omega \exp\left( - i\omega t\right)\right.  \nonumber\\ ~&+~ \left. Y_k^{\omega *} \exp\left( i\omega t\right) \right\rbrace 
+~ \sum_{k}A_{n k}^{*}\left\lbrace X_k^{\omega *} \exp\left( i\omega t \right)~+~ Y_k^\omega \exp\left( -i\omega t \right)\right\rbrace
\label{cc_rpasol}
\end{align}
Since \(\exp(-i\omega t)\) and \(\exp(i\omega t)\) represent linearly independent solutions, Eqs.~\eqref{coupled} and \eqref{cc_rpasol} can be separated accordingly. This leads to a matrix representation of the system, corresponding to the Second Random Phase Approximation (SRPA), which includes not only residual interactions within each sector individually but also the coupling between the \(1p\text{--}1h\) (\(1\alpha\text{--}1h\)) and \(2p\text{--}2h\) sectors,
\begin{equation}
\begin{pmatrix}
A & B \\
-B^{*} & -A^{*}
\end{pmatrix}
\begin{pmatrix}
X \\
Y
\end{pmatrix}
=
\hbar \omega
\begin{pmatrix}
X \\
Y
\end{pmatrix},
\label{eq:srpa_matrix}
\end{equation}
where $A$ and $B$ matrix contain the following blocks:
\begin{equation}
A = \begin{pmatrix}
A_{11} & A_{12} \\
A_{21} & A_{22}
\end{pmatrix},
\quad \text{and} \quad
B = \begin{pmatrix}
B_{11} & B_{12} \\
B_{21} & B_{22},
\end{pmatrix}
\end{equation}
and each of them can be written in the form:
\begin{equation}
\begin{aligned}
A_{11} =  \begin{pmatrix}
A_{ph;p^\prime h^\prime} & A_{ph;\alpha^\prime h^\prime} \\
A_{\alpha h;\alpha^\prime h^\prime} & A_{\alpha h; \alpha^\prime h^\prime}
\end{pmatrix},
\quad
A_{12} =  \begin{pmatrix}
A_{ph;p_1^\prime p_2^\prime h_1^\prime h_2^\prime} \\
A_{\alpha h;p_1^\prime p_2^\prime h_1^\prime h_2^\prime}
\end{pmatrix},
\quad
A_{22} =A_{p_1p_2h_1h_2;p_1^\prime p_2^\prime h_1^\prime h_2^\prime}
\\
B_{11} =  \begin{pmatrix}
B_{ph \overline{p^\prime h^\prime};0} & B_{p h\overline{\alpha^\prime h^\prime};0} \\
B_{\alpha h \overline{\alpha^\prime h^\prime};0} & B_{\alpha h \overline{\alpha^\prime h^\prime};0}
\end{pmatrix},
\quad
B_{12} = \begin{pmatrix} 
B_{ph\overline{p_1^\prime p_2^\prime h_1^\prime h_2^\prime};0} \\
B_{\alpha h \overline{p_1^\prime p_2^\prime h_1^\prime  h_2^\prime};0} 
\end{pmatrix},
\quad
B_{22} = 0
\end{aligned}
\end{equation}
This matrix structure follows the conventional notation introduced in the context of the RPA and its extensions~\cite{Rowe1970}.
The states labeled by $\overline{p_1h_1}$ and $\overline{p_1p_2h_1h_2}$ correspond to the R-conjugate $p$--$h$ and $2p$--$2h$ configurations defined through time-reversal symmetry. In the spherical case, this symmetry implies that conjugate configurations differ only by the sign of the spin projection quantum number, with all other quantum numbers remaining unchanged~\cite{Rowe1970}.
Here, \(A_{11}\) and \(B_{11}\) are the standard RPA (sub)matrices constructed from $p$--$h$ and $\alpha$--$h$ configurations, and arise from the variation with respect to the amplitudes \(Z^{(1)}\) and \(Z^{(1)*}\). The coupling blocks \(A_{12}\) and \(A_{21} = A_{12}^\dagger\) encode the interaction between \(1p\text{--}1h\) (\(1\alpha\text{--}1h\)) and \(2p\text{--}2h\) configurations and are responsible for the fragmentation and damping of collective modes beyond the RPA level. 
In addition to the standard matrix blocks, the term \(B_{12}\) appears exclusively in the case of energy density functional (EDF) theories, where the underlying interaction exhibits an explicit dependence on the density (or four-current). This matrix block contains only rearrangement terms arising from the functional derivatives of the density-dependent couplings and has no counterpart in interaction models based on fixed two-body forces. The $A_{22}$ block contains the interaction between two-particle--two-hole ($2p$--$2h$) configurations, which plays a central role in shaping the fragmentation and spreading of the nuclear excitation spectrum.

The explicit forms of these matrix elements depend on the details of the effective interaction and the structure of the underlying relativistic single-particle basis. The dimension of the matrix increases rapidly with the number of \(2p\text{--}2h\) configurations, making truncation schemes and efficient computational strategies essential for practical implementations.

\subsection{Relativistic density matrix}\label{sec:relden}
Using the fact that there exists a connection between the ground state and an arbitrary excited state of the nucleus through expression~\eqref{eq:full_ansatz}, it is possible to write the one-body density matrix as \cite{Gambacurta2015}
\begin{equation}
\begin{aligned}
\rho_{a b} & \simeq \langle\Phi_0| e^{\hat{Z}^{\dagger}} a_b^{\dagger} a_a e^{\hat{Z}}|\Phi_0\rangle \\
& =\langle\Phi_0|\left(1+\hat{Z}^{\dagger}+\frac{1}{2}\left(\hat{Z}^{\dagger}\right)^2+\ldots\right) a_b^{\dagger} a_a\left(1+\hat{Z}+\frac{1}{2}(\hat{Z})^2+\ldots\right)|\Phi_0\rangle .
\end{aligned}
\end{equation}
Expanded in contributions of the Thouless operator \( \hat{Z} \) up to second order, the one-body density matrix then takes the form
\begin{equation}
\begin{aligned}
\rho_{a b}~=&~\langle\Phi_0| a_b^{\dagger} a_a|\Phi_0\rangle+\langle\Phi_0| \hat{Z}^{\dagger} a_b^{\dagger} a_a+a_b^{\dagger} a_a \hat{Z}|\Phi_0\rangle\\
&+~\langle\Phi_0| \frac{1}{2}\left(\hat{Z}^{\dagger}\right)^2 a_b^{\dagger} a_a
+\hat{Z}^{\dagger} a_b^{\dagger} a_a \hat{Z}+\frac{1}{2} a_b^{\dagger} a_a \hat{Z}^2|\Phi_0\rangle .
\end{aligned}
\label{Zexpand}
\end{equation}
The first term presents zero-th order of the density matrix, i.e., \( \rho^{(0)}_{ab} \), which is nothing but the 
\begin{equation}
\rho_{a b}^{(0)}=\delta_{a b} \theta\left(\epsilon_F-\epsilon_a\right)\quad\rightarrow\quad \rho_{h h^\prime}^{(0)} = \delta_{h h^\prime},
\end{equation}
where $\theta$ is the Heaviside step function, indicating fully occupied hole states below the Fermi energy and empty particle states above it. Clearly, at zeroth order, only the diagonal elements below the Fermi level $\epsilon_F$ contribute. All other combinations which include $p\text{-}h$, $\alpha\text{--}h$, but also $p\text{--}p$, $p\text{--}\alpha$ and $\alpha\text{--}\alpha$ pairs vanish identically.

The linear contribution of the Thouless operator to the matrix element of the one-body density is nothing but the second term in expression \eqref{Zexpand}, i.e., the corresponding bra-ket:
\begin{equation}
\begin{aligned}
\rho_{a b}^{(1,1)} & =\sum_{m i}\langle\Phi_0| Z_{m i}^{(1)*} a_i^{\dagger} a_m a_b^{\dagger} a_a+a_b^{\dagger} a_a Z_{m i}^{(1)} a_m^{\dagger} a_i|\Phi_0\rangle \\
\rho_{a b}^{(1,2)} & =\frac{1}{2} \sum_{m n i j}\langle\Phi_0| Z_{m n i j}^{(2)*} a_i^{\dagger} a_j^{\dagger} a_n a_m a_b^{\dagger} a_a+a_b^{\dagger} a_a Z_{m n i j}^{(2)} a_m^{\dagger} a_n^{\dagger} a_j a_i|\Phi_0\rangle
\end{aligned}
\label{one_and_two}
\end{equation}
which leads to
\begin{equation}
\begin{aligned}
& \delta \rho_{h p}^{(1)}=Z_{p h}^{(1)*},\quad\delta \rho_{h \alpha}^{(1)}=Z_{\alpha h}^{(1)*} \\
& \delta \rho_{p h}^{(1)}=Z_{p h}^{(1)},\quad\delta \rho_{\alpha h}^{(1)}=Z_{\alpha h}^{(1)}
\end{aligned}
\end{equation}
The contribution of the $2p\text{--}2h$ configurations (the second bra-ket in Eq.~\eqref{one_and_two}) vanishes because, after the canonical transformation of the operators, there are always at least two particle (hole) creation operators, at least one of which can never be matched with a corresponding particle (hole) annihilation operator. In other words, $2p\text{--}2h$ configurations do not contribute at first order to \( \rho^{(1)}_{ab} \). Clearly, we can state that
\begin{equation}
\begin{aligned}
& \delta \rho_{p p^{\prime}}^{(1)}=\delta \rho_{\alpha p^{\prime}}^{(1)} = \delta \rho_{p \alpha^{\prime}}^{(1)} = \delta \rho_{\alpha p^{\prime}}^{(1)}= 0 \\
& \delta \rho_{h h^{\prime}}^{(1)}=0
\end{aligned}
\end{equation}
Similar results for nonrelativistic case are obtained in \cite{Gambacurta2015}.
The quadratic contribution in \( \hat{Z} \) can be expressed by the following bra-ket:
\begin{equation}
\langle\Phi_0| \frac{1}{2} a_b^{\dagger} a_a \hat{Z}^2+\hat{Z}^{\dagger} a_b^{\dagger} a_a \hat{Z}+\frac{1}{2}\left(\hat{Z}^{\dagger}\right)^2 a_b^{\dagger} a_a|\Phi_0\rangle.
\label{raspis_second_order}
\end{equation}
The only nonvanishing contribution comes from the second term in Eq. \eqref{raspis_second_order}, from which follow second order densities within the {$1p$--$1h$} $\oplus$ {$1\alpha$--$1h$} $\oplus$ {$2p$--$2h$} space: 
\begin{equation}
\begin{aligned}
\delta \rho_{p h}^{(2)}= & \sum_{m i} Z_{m i}^{(1)*} Z_{m p i h}^{(2)},\quad \delta \rho_{h p}^{(2)}= \sum_{m i} Z_{m i}^{(1)} Z_{m p i h}^{(2)*},  \\
\delta \rho_{p p^{\prime}}^{(2)}= & \sum_i Z_{p i}^{(1)*} Z_{p^{\prime} i}^{(1)}+\frac{1}{2} \sum_{m i \neq j} Z_{m p i j}^{(2)*} {Z}_{m p^{\prime} i j}^{(2)}, \\
\delta \rho_{\alpha \alpha^{\prime}}^{(2)}= & \sum_i Z_{\alpha i}^{(1)*} Z_{\alpha^{\prime} i}^{(1)}, \quad \delta \rho_{\alpha h^{\prime}}^{(2)} =\delta \rho_{h \alpha^{\prime}}^{(2)} = 0\\ 
\delta \rho_{\alpha p^{\prime}}^{(2)}= & \sum_i Z_{\alpha i}^{(1)*} Z_{p^{\prime} i}^{(1)}, \quad \delta \rho_{p \alpha^{\prime}}^{(2)}=  \sum_i Z_{p i}^{(1)*} Z_{\alpha^{\prime} i}^{(1)},\\
\delta \rho_{h h^{\prime}}^{(2)}= &  -\left.\sum_{m } Z_{m h}^{(1)*} Z_{m h^\prime}^{(1)} - \frac{1}{2} \sum_{m \neq n}\sum_{~j}{ } {Z}_{m n h j}^{(2)*} {Z}_{m n h^\prime j}^{(2)}\right. .
\end{aligned}
\end{equation}

\subsection{Explicit expressions for the RSRPA matrix elements}\label{sec:matrix_structure}

The RSRPA matrix elements are derived from the second functional derivative of the energy of a system in arbitrary state with respect to the amplitudes \(Z^{(1)}_{ph}(t)\), \(Z^{(1)}_{\alpha h}(t)\), \(Z^{(2)}_{pp'hh'}(t)\) and c.c. In the relativistic framework, the following derivation is general and holds regardless of whether one considers effective interaction based on meson exchange or a point-coupling interaction constructed from densities and currents bilinear in the Dirac spinor field.
The explicit expressions for the RSRPA matrix blocks \(A_{11}\), \(A_{12}\), and \(A_{22}\) are given below.

The \(A_{11}\) block corresponds to the standard TDA matrix or the A submatrix in the (R)RPA case, involving particle–hole (\(p\text{--}h\)) excitations, which are obtained from:
\begin{equation}
\left[A_{11}\right]_{ph;p^\prime h^\prime}~=~\left.\frac{\partial^2\langle \Phi(t) | \hat{H} |\Phi(t)\rangle}{\partial Z_{p^\prime h^\prime} \partial Z_{p h}^{*}}\right\vert_{Z_{ph}, ~{Z}_{p_1 p_2 h_1 h_2}\text{ and c. c.}~=~0},
\label{A11phph}
\end{equation}
with the expansion of the two-body residual interaction $\hat{v}_{\alpha\beta\gamma\delta}$ up to the second order in density matrix:
\begin{equation}
\hat{v}_{\alpha\beta\gamma\delta}~=~\hat{v}_{\alpha\beta\gamma\delta}\left[ \rho_0 \right]~+~\sum_{a b}\left.\frac{\partial \hat{v}_{\alpha\beta\gamma\delta}}{\partial \rho_{ab}}\right\vert_{\rho=\rho^{(0)}} d\rho_{ab}~+~\left.\frac{1}{2}\sum_{abcd}\frac{\partial^2 \hat{v}_{\alpha\beta\gamma\delta}}{\partial\rho_{ab}\partial\rho_{cd}}\right\vert_{\rho~=~\rho^{(0)}}d\rho_{ab}d\rho_{cd},
\label{residualexp}
\end{equation}
and 
\begin{equation}
d\rho_{ab}~=~\delta\rho_{ab}^{(1)}+\delta\rho_{ab}^{(2)}.
\label{rho12}
\end{equation}
The indices \(a\), \(b\), \(c\), and \(d\), as well as \(\alpha\), \(\beta\), \(\gamma\), and \(\delta\) in Eqs. \eqref{residualexp} and \eqref{rho12} are in general associated with particle, hole, or antiparticle states. After subtitution of Eqs. \eqref{residualexp} and \eqref{rho12} in Eq. \eqref{A11phph} for corresponding $A_{11}$ block we have:
\begin{equation}
\begin{aligned}
\relax [A_{11}]_{ph,p'h'} ~=&~ (\epsilon_p - \epsilon_h)\, \delta_{pp'} \delta_{hh'}
+ \hat{v}_{p h' h p'} ~+~\sum_j\left.\left[\frac{\partial \hat{v}_{pjhj}\left[\rho\right]}{\partial \rho_{p'h'}}\right]\right\vert_{\rho~=~\rho^{(0)}} \\~&+~\sum_j\left[\frac{\partial \hat{v}_{{h'}jp'j}\left[\rho\right]}{\partial \rho_{hp}} \bigg]\right\vert_{\rho=\rho^{(0)}}
+~\frac{1}{2} \sum_{ij}\left.\bigg[\frac{\partial^2 \hat{v}_{ijij}\left[\rho\right]}{\partial\rho_{hp}\partial\rho_{{p'}h'}}\bigg]\right\vert_{\rho~=~\rho^{(0)}}
\end{aligned}
\label{eq:A11}
\end{equation}
where \(\epsilon_i\) are the Dirac single-particle energies. 
Any term that, after taking the second derivative with respect to the expansion coefficients, and after doing the expansion of the two-body interaction matrix element, still contains any of the expansion coefficients, vanishes identically. One of the \( A_{11} \) subblocks, representing the interaction between \(1p\text{--}1h\) and \(1\alpha\text{--}1h\) configurations, is obtained from the variation of the action functional with respect to \( Z_{p'h'} \) and \( Z^*_{\alpha h} \), i.e.,
\begin{equation}
\left[A_{11}\right]_{\alpha h;p'h'}\frac{\partial^2\langle \Phi(t) | \hat{H} |\Phi(t)\rangle}{\partial Z_{p^\prime h^\prime} \, \partial Z_{\alpha h}^{*}},
\end{equation}
which, when expressed as a functional derivative of the two-body interaction matrix element with respect to the density matrix, takes the following form,
\begin{equation}
\begin{aligned}
\relax [A_{11}]_{\alpha h,p'h'} ~=&~ \hat{v}_{\alpha h' h p'} ~+~\sum_j\left.\left[\frac{\partial \hat{v}_{\alpha jhj}\left[\rho\right]}{\partial \rho_{p'h'}}\right]\right\vert_{\rho~=~\rho^{(0)}} ~+~\sum_j\left[\frac{\partial \hat{v}_{{h'}jp'j}\left[\rho\right]}{\partial \rho_{h \alpha}} \bigg]\right\vert_{\rho=\rho^{(0)}}\\
&+~\frac{1}{2} \sum_{ij}\left.\bigg[\frac{\partial^2 \hat{v}_{ijij}\left[\rho\right]}{\partial\rho_{h\alpha}\partial\rho_{{p'}h'}}\bigg]\right\vert_{\rho~=~\rho^{(0)}}.
\end{aligned}
\label{eq:A11alfap}
\end{equation}
%
Analogous expressions are obtained for the remaining two subblocks of the \(A_{11}\) matrix by performing the appropriate index substitutions. 

The coupling block \(A_{12}\) in the relativistic framework connects the \(1p\text{--}1h\) or \(1\alpha\text{--}1h\), with \(2p\text{--}2h\) configuration space. In the first case, we have the following variation with respect to the \( Z \) coefficients,
\begin{equation}
\begin{aligned}
\left[A_{12}\right]_{ph;p_1^\prime p_2^\prime h_1^\prime h_2^\prime}~=&~ \left.\frac{\partial^2\langle \Phi(t) | \hat{H} |\Phi(t)\rangle}{\partial Z_{ph}^{*} \partial Z_{p_1^\prime p_2^\prime h_1^\prime h_2^\prime}}\right\vert_{Z_{mi}, ~{Z}_{mnij}\text{ and c. c.}~=~0},
\end{aligned}
\end{equation}
which can be written as
\begin{equation}
\begin{aligned}
\left[A_{12}\right]_{{p}h;p_1 p_2 h_1h_2}~=&~ \mathcal{A}(h_1,h_2) \delta_{h h_1} \hat{v}_{{p} h_2 p_1 p_2} \left[ \rho^{(0)}\right]~-~\mathcal{A}(p_1,p_2) \delta_{{p} p_1} \hat{v}_{h_1 h_2 h p_2} \left[ \rho^{(0)}\right]\\
~&+~\left.\frac{\partial \hat{v}_{h_1h_2p_1 p_2}\left[ \rho\right]}{\partial \rho_{hp}}\right\vert_{\rho=\rho^{(0)}}.
\end{aligned} 
\label{A12standard}
\end{equation}
The $A_{12}$ block describing the coupling between \(1\alpha\)–\(1h\) and \(2p\)–\(2h\) configurations is given with the following variation,
\begin{equation}
\begin{aligned}
\left[A_{12}\right]_{\alpha h;p_1^\prime p_2^\prime h_1^\prime h_2^\prime}~=&~ \left.\frac{\partial^2\langle \Phi(t) | \hat{H} |\Phi(t)\rangle}{\partial Z_{\alpha h}^{*} \partial Z_{p_1^\prime p_2^\prime h_1^\prime h_2^\prime}}\right\vert_{Z_{mi}, ~{Z}_{mnij}\text{ and c. c.}~=~0},
\end{aligned}
\end{equation}
which can be further written as
\begin{equation}
\begin{aligned}
\left[A_{12}\right]_{{\alpha}h;p_1 p_2 h_1h_2}~=&~ \mathcal{A}(h_1,h_2) \delta_{h h_1} \hat{v}_{{\alpha} h_2 p_1 p_2} \left[ \rho^{(0)}\right] + \left.\frac{\partial \hat{v}_{h_1h_2p_1 p_2}\left[ \rho\right]}{\partial \rho_{h\alpha}}\right\vert_{\rho=\rho^{(0)}}.
\end{aligned} 
\label{A12standard2}
\end{equation}

The \(A_{22}\) block contains matrix elements between \(2p\text{--}2h\) configuration given as,
\begin{equation}
\begin{aligned}
\left[A_{22}\right]_{p_1h_1p_2h_2;p_1^\prime p_2^\prime h_1^\prime h_2^\prime}~=&~ \left.\frac{\partial^2\langle \Phi(t) | \hat{H} |\Phi(t)\rangle}{\partial Z_{p_1p_2h_1h_2}^{*} \partial Z_{p_1^\prime p_2^\prime h_1^\prime h_2^\prime}}\right\vert_{Z_{mi}, ~{Z}_{mnij}\text{ and c. c.}~=~0},
\end{aligned}
\end{equation}
which takes the following form
\begin{equation}
\begin{aligned}
\relax [A_{22}]_{p_1p_2h_1h_2;{p_1^{\prime}}{p_2^{\prime}}{h_1^{\prime}}{h_2^{\prime}}}~&=~\delta_{p_1 p_1^\prime} \delta_{p_2 p_2^\prime} \delta_{h_1 h_1^\prime} \delta_{h_2 h_2^\prime}\, ( \epsilon_{p_1} + \epsilon_{p_2} - \epsilon_{h_1} - \epsilon_{h_2} )\\
~&~~~+~\mathcal{A}(h_1,h_2) \delta_{h_1{h_1^{\prime}}}\delta_{h_2{h_2^{\prime}}}\hat{v}_{p_1p_2{p_1^{\prime}}{p_2^{\prime}}}~+~\mathcal{A}(p_1,p_2)\delta_{p_1{p_1^{\prime}}}\delta_{p_2{p_2^{\prime}}}\hat{v}_{{h_1^{\prime}}{h_2^{\prime}}h_1h_2}\\
~&~~~+~\mathcal{A}(p_1,p_2)\mathcal{A}(h_1,h_2)\mathcal{A}({p_1^{\prime}},{p_2^{\prime}})\mathcal{A}({h_1^{\prime}},{h_2^{\prime}})\delta_{p_1{p_1^{\prime}}}\delta_{h_2{h_2^{\prime}}}\hat{v}_{p_2{h_1^{\prime}}{p_2^{\prime}}h_1} \\
~&~~~+~\left.\frac{1}{2}\sum_{hh'}\mathcal{A}(p_1,p_2) \mathcal{A}(p_1^\prime,p_2^\prime)\frac{\partial\hat{v}_{hh'hh'}\left[\rho\right]}{\partial\rho_{p_2^\prime p_2}}\right\vert_{\rho=\rho^{(0)}}\delta_{p_1 p_1^\prime}\delta_{h_1 h_1^\prime}\delta_{h_2 h_2^\prime}\\
&~~~-~\left.\frac{1}{2}\sum_{hh'}\mathcal{A}(h_1,h_2) \mathcal{A}(h_1^\prime,h_2^\prime)\frac{\partial\hat{v}_{hh'hh'}\left[\rho\right]}{\partial\rho_{h_1 h_1^\prime}}\right\vert_{\rho=\rho^{(0)}}\delta_{p_1 p_1^\prime}\delta_{p_2 p_2^\prime}\delta_{h_2 h_2^\prime}.
\end{aligned}
\end{equation}
The diagonal part accounts for the unperturbed \(2p\text{--}2h\) excitation energies, while the off-diagonal terms reflect interaction-induced couplings among complex configurations. Furthermore, if the residual interaction among the two-particle--two-hole (\(2p\text{--}2h\)) configurations is neglected, the matrix \(A_{22}\) reduces to a diagonal form, with elements corresponding to the unperturbed \(2p\text{--}2h\) excitation energies,
\begin{equation}
[A_{22}^{\text{diag}}]_{p_1p_2h_1h_2;{p_1^{\prime}}{p_2^{\prime}}{h_1^{\prime}}{h_2^{\prime}}} = \delta_{p_1 p_1^{\prime}}\, \delta_{h_1 h_1^{\prime}}\, \delta_{p_2 p_2^{\prime}}\, \delta_{h_2 h_2^{\prime}}\, \left(\epsilon_{p_1} + \epsilon_{p_2} - \epsilon_{h_1} - \epsilon_{h_2} \right).
\label{diagonal}
\end{equation}
This approximation defines the diagonal RSRPA, hereafter denoted as RSRPA(d). In the standard RPA, the matrix \(B\) contains the coupling between backward-going amplitudes particle–hole pairs and plays a crucial role in ensuring the correct symmetry of the RPA equations and the stability of the ground state. In the RSRPA formalism, the \(B\) matrix is generalized to include additional blocks involving the coupling of \(1p\text{--}1h\) and \(2p\text{--}2h\) configurations, where the latter is arising from the rearrangement contributions in density-dependent interactions. 
\noindent
The standard $B_{11}$ block is obtained
\begin{equation}
\left[B_{11}^{*}\right]_{0;p_1 h_1\overline{p_2 h_2}}=\left.\frac{\partial^2\langle \Phi(t) | \hat{H} |\Phi(t)\rangle}{\partial Z_{p_1 h_1} \partial Z_{p_2 h_2}}\right|_{Z_{m i}, \hat{Z}_{m n i j} \text { and c. c. }~=~0}.
\end{equation}
from which follows
\begin{equation}
\begin{aligned}
{\left[B_{11}^{*}\right]_{0 ; p_1 h_1 \overline{p_2 h_2}} }  = &
\left.\frac{1}{2} \sum_{h h^\prime} \left[\frac{\partial^2 \hat{v}_{h h^\prime h h^\prime}[\rho]}{\partial \rho_{p_1 h_1} \partial \rho_{p_2 h_2}}\right]\right|_{\rho=\rho^{(0)}}
+\left.\sum_h\left[\frac{\partial \hat{v}_{h_1 h p_1 h} [\rho]}{\partial \rho_{p_2 h_2}}\right]\right|_{\rho=\rho^{(0)}} \\
&+\left.
\sum_h \left[ \frac{\partial \hat{v}_{h_2 h p_2 h} [\rho]}{\partial \rho_{p_1 h_1}}\right]\right|_{\rho=\rho^{(0)}} 
 +\hat{v}_{h_2 h_1 p_2 p_1} [\rho]
\end{aligned}
\end{equation}
Similar expressions are obtained for $B_{11}$ subblocks which involve correlations between $p\text{--}h$ and $\alpha\text{--}h$ pairs, as well as between two $\alpha\text{--}h$ pairs.

For the $B_{12}$, which involve the correlations between $p\text{--}h$ and $2p\text{--}2h$ configurations, we have
\begin{equation}
\left[B_{12}^{*}\right]_{0; p h\overline{p_1 p_2 h_1 h_2} }=\left.\frac{\partial^2\langle \Phi(t) | \hat{H} |\Phi(t)\rangle}{\partial Z_{p h} \partial {Z}_{p_1 p_2 h_1 h_2}}\right|_{Z_{m i},{Z}_{m n i j} \text { and c. c. }=~0}
\label{B12gen}
\end{equation}
By applying the expression \eqref{B12gen} we obtain
\begin{equation}
\begin{aligned}
\left[B_{12}^{*}\right]_{0 ;p h\overline{p_1 p_2 h_1 h_2}} & =\left.\left[\frac{\partial \hat{v}_{p_1 p_2 h_1 h_2}[\rho]}{\partial\rho_{ph}}\right]\right|_{\rho=~\rho^{(0)}}
\end{aligned}
\end{equation}
An entirely analogous expression is obtained in the case of correlations involving $1\alpha\text{--}1h$ and $2p\text{--}2h$ configurations, with proper substitution of indices $p\rightarrow\alpha$. The block $B_{12}$ consists solely of rearrangement terms, i.e., it is nonzero only for density dependent interactions.

Up to this point, all derivations remain fully valid for both classes of relativistic energy density functionals, i.e., those based on point-coupling and meson-exchange interactions. Moreover, the formalism is applicable regardless of whether the ground state is obtained within the relativistic Hartree (RH) or relativistic Hartree–Fock (RHF) approximation, provided that the residual interaction is derived consistently from the underlying Lagrangian density. A detailed derivation of the coupled matrix elements \(A_{11}^J\), \(B_{11}^J\), \(A_{12}^J\), and \(A_{22}^J\) within the SRPA formalism is provided in Ref.~\cite{Papakonstantinou2010}. 

\subsection{Point-coupling interaction}\label{sec:interaction}

In the RSRPA formulation in this work we employ relativistic density-dependent point coupling interaction~\cite{Niksic2008}.
Upon including the free-nucleon and electromagnetic components, the resulting Lagrangian density for the nuclear system is given by the expression \cite{Nik}:
\begin{equation}
\begin{aligned}
\mathcal{L}~=&~\bar{\psi}\left(i\gamma_\mu\partial^\mu - m \right)\psi-\frac{1}{2}\alpha_S(\rho)\left(\bar{\psi}\psi\right)\left(\bar{\psi}\psi\right)\\
&-\frac{1}{2}\alpha_V(\rho)\left(\bar{\psi}\gamma_\mu\psi\right)\left(\bar{\psi}\gamma^\mu\psi\right)-\frac{1}{2}\alpha_{TV}(\rho)\left(\bar{\psi}\vec{\tau}\gamma_\mu\psi\right)\left(\bar{\psi}\vec{\tau}\gamma^\mu\psi\right)
\\
&-\frac{1}{2}\delta_S\partial_\nu\left(\bar{\psi}\psi\right)\partial^\nu\left(\bar{\psi}\psi\right)
-e\bar{\psi}\gamma_\mu A^\mu \frac{1-\tau_3}{2}\psi.
\label{lag}
\end{aligned}
\end{equation}
The density dependence is explicitly included in the isoscalar-scalar $\alpha_S(\rho)$, isoscalar-vector $\alpha_V(\rho)$ and isovector-vector couplings $\alpha_{TV}(\rho)$ as 
\begin{equation}
\alpha_i\left[ x \right]~=~a_i + \left( b_i + c_i x\right)\exp(-d_i x),
\end{equation}
where $x = \rho/\rho_{sat}$, $\rho$ stands for barion (vector) density and $\rho_{sat}$ for the nucleon density at saturation in the symmetric nuclear matter \cite{Nik}. In the calculations, we employ the DD-PCX parameterization for $a_i, ~b_i, ~c_i$ and $d_i$ as given in Ref. \cite{Yuk2019}. The DD-PCX parameterization was introduced as a new relativistic density-dependent point-coupling interaction, optimized by simultaneously fitting both ground-state properties and experimental data on collective excitations, including the isoscalar giant monopole resonance and dipole polarizability in \(^{208}\)Pb~\cite{Yuk2019}. 
The formalism for computing relativistic two-body matrix elements, both in coupled and uncoupled representations for specific particle–hole channels, is given in Ref.~\cite{Vale2}. By analogy, the same methodology from Ref.~\cite{Vale2} can be extended to other particle--hole channels by applying the corresponding angular momentum algebra.

\subsection{Strength function and norm of the excited states} \label{sec:transitions}

To probe collective spin independent modes within the RSRPA framework, we introduce external field operators,
given by expression:
\begin{equation}
\begin{aligned}
\hat{F}_{\lambda M}^{\text{IS}} ~=&~ \sum_i r_i^k Y_{\lambda M} (\Omega_i),   \\
\hat{F}_{\lambda M}^{\text{IV}} ~=&~ \sum_i r_i^k Y_{\lambda M} (\Omega_i) \hat{t}_z(i),
\end{aligned}    
\end{equation}
for the isoscalar (IS) and isovector (IV) channel, respectively. For $\lambda = 0$, i.e. the monopole case, $k = 2$, while for other cases $k = \lambda$. 
The strength distribution associated with this operator is defined as
\begin{equation}
\begin{aligned}
R(E) ~=&~ \sum_\nu |\langle \nu | \hat{F} | 0 \rangle|^2\, \delta(E - E_\nu),\\
=&~\sum_\nu B_F(E_\nu)\delta(E - E_\nu),
\label{eq:strength_function}
\end{aligned}
\end{equation}
where \(|\nu\rangle\) are the excited states obtained by solving the SRPA eigenvalue problem, and \(E_\nu\) are the corresponding excitation energies.

The energy weighted moments of the transition strength distribution are defined as
\begin{equation}
m_k = \int E^k R(E)\, dE,
\label{moments}
\end{equation}
where \( k \) represents the order of the moment. Based on these quantities, the centroid energy, which characterizes the mean energy of the distribution, is expressed as
\begin{equation}
\bar{E} = \frac{m_1}{m_0}.
\end{equation}
The moments \( m_k \) can be computed over the entire excitation spectrum, over specific parts such as the positive energy domain, or within a selected experimental interval. The width of the resonance, which quantifies the spread of the strength around the centroid, is calculated as
\begin{equation}
\Delta = \sqrt{ \frac{m_2}{m_0} - \bar{E}^2 }.
\end{equation}

Within the present SRPA formalism, the excited state \(|\nu\rangle\) is represented as a linear combination of particle–hole, antiparticle-hole and two-particle–two-hole configurations:
\begin{equation}
\begin{aligned}
    |\nu\rangle =& \sum_{ph} \left( X^{(1)\nu}_{ph} a_p^\dagger a_h - Y^{(1)\nu}_{ph} a_h^\dagger a_p \right) |\Phi_0\rangle + \sum_{\alpha h} \left( X^{(1)\nu}_{\alpha h} a_\alpha^\dagger a_h - Y^{(1)\nu}_{\alpha h} a_h^\dagger a_\alpha \right) |\Phi_0\rangle \\
&+ \frac{1}{4} \sum_{pp'hh'} \left( X^{(2)\nu}_{pp'hh'} a_p^\dagger a_{p'}^\dagger a_{h'} a_h 
- Y^{(2)\nu}_{pp'hh'} a_h^\dagger a_{h'}^\dagger a_{p'} a_p \right) |\Phi_0\rangle.
\label{eq:SRPA_state}
\end{aligned}
\end{equation}
The transition matrix element is then computed as
\begin{equation}
\langle \nu | \hat{F} | 0 \rangle = \sum_{ph} \left( f_{ph}\, X^{(1)\nu}_{ph} + f_{hp}\, Y^{(1)\nu}_{ph} \right) + \sum_{\alpha h} \left( f_{\alpha h}\, X^{(1)\nu}_{\alpha h} + f_{h\alpha}\, Y^{(1)\nu}_{\alpha h} \right),
\label{eq:transition_matrix}
\end{equation}
where \(f_{ph} = \langle p | \hat{F} | h \rangle\) are the single-particle transition matrix elements of the operator \(\hat{F}\). Contributions from \(2p\text{--}2h\) configurations vanish at this order because \(\hat{F}\) is a one-body operator and does not connect the reference state \(|\Phi_0\rangle\) directly to \(2p\text{--}2h\) states.
Although $2p$--$2h$ configurations do not contribute directly to the transition matrix element, they mix with $1p$--$1h$ configurations through the SRPA equations. Accordingly, the amplitudes $X^{(1)\nu}$ and $Y^{(1)\nu}$ used in Eq.(\ref{eq:transition_matrix}) are influenced by coupling to  $2p$--$2h$ configuration space.

The normalization condition for the RSRPA amplitudes reflects the extended quasiboson structure of the excited state. Assuming that the creation operator \(Q_\nu^\dagger\) of the mode \(\nu\) satisfies \(Q_\nu |0\rangle = 0\), the norm is given by 
\begin{equation}
\mathcal{N}_\nu  ~=~   \mathcal{N}_1  + \mathcal{N}_2 ~=~ 1
\label{eq:srpa_norm}
\end{equation}
where
\begin{equation}
\begin{aligned}
\mathcal{N}_1 ~=&~ \sum_{ph} \left( |X^{(1)\nu}_{ph}|^2 - |Y^{(1)\nu}_{ph}|^2 \right) 
+ \sum_{\alpha h} \left( |X^{(1)\nu}_{\alpha h}|^2 - |Y^{(1)\nu}_{\alpha h}|^2 \right),\\
\mathcal{N}_2 ~=&~ \sum_{pp'hh'} \left( |X^{(2)\nu}_{pp'hh'}|^2 - |Y^{(2)\nu}_{pp'hh'}|^2 \right)\\
\end{aligned}
\label{eq:srpa_norm_part}
\end{equation}
Eqs. \eqref{eq:srpa_norm} and \eqref{eq:srpa_norm_part} serve as a useful diagnostic of the collectivity and internal consistency of the excitation spectrum, particularly in situations where the full SRPA matrix is truncated, for example by imposing an energy cutoff on the \(2p\text{--}2h\) configuration space, by using the diagonal approximation for the \(A_{22}\) submatrix, or by solving only a restricted part of the eigenvalue problem.

\subsection{Subtraction method}\label{sec:subtraction}
The standard formulation of the SRPA includes explicit two-particle–two-hole (\(2p\text{--}2h\)) configurations, whose coupling to the particle–hole (\(1p\text{--}1h\)) space induces a notable downward shift of excitation energies relative to the conventional RPA \cite{Papakonstantinou2010, Gambacurta2012,Gabacurta2009}. To resolve this issue and restore consistency with the static RPA limit, Tselyaev~\cite{Tselyaev2007} introduced the subtraction method, whereby the static part of the energy-dependent effective interaction is removed,
\begin{equation}
V^{\text{SRPA}}(\omega) \rightarrow V^{\text{SRPA}}(\omega) - [V^{\text{SRPA}}(0) - V^{\text{RPA}}].
\end{equation}
This procedure ensures that the extended SRPA response function reduces to the RPA result in the zero-frequency limit, thereby preventing double-counting of static correlations and improving the stability of the excitation spectrum.

The first implementation of the subtraction method within the framework of non-relativistic SRPA was carried out by Gambacurta \textit{et al.}~\cite{Gambacurta2015}, whose formulation we adopt in the present relativistic extension. Their scheme provides a consistent prescription for removing the static contribution of the energy-dependent interaction, which for $A$ and $B$ matrices takes the following form:
\begin{equation}
\begin{aligned}
\mathcal{A}_F^S & =\left(\begin{array}{cc}
A_{11^{\prime}}+\sum_{2,2^{\prime}} A_{12}\left(A_{22^{\prime}}\right)^{-1} A_{2^{\prime} 1^{\prime}}+\sum_{2,2^{\prime}} B_{12}\left(A_{22^{\prime}}\right)^{-1} B_{2^{\prime} 1^{\prime}} & A_{12} \\
A_{21} & A_{22^{\prime}}
\end{array}\right), \\
\mathcal{B}_F^S & =\left(\begin{array}{cc}
B_{11^{\prime}}+\sum_{2,2^{\prime}} A_{12}\left(A_{22^{\prime}}\right)^{-1} B_{2^{\prime} 1^{\prime}}+\sum_{2,2^{\prime}} B_{12}\left(A_{22^{\prime}}\right)^{-1} A_{2^{\prime} 1^{\prime}} & B_{12} \\
B_{21} & 0
\end{array}\right).
\end{aligned}
\end{equation}
The above structure represents the subtracted formulation of the RSRPA, denoted as RSSRPA. For the diagonal case of the subtracted RSRPA (RSSPRA(d)) the expressions reduce to 
\begin{equation}
\begin{gathered}
\mathcal{A}_D^S=\left(\begin{array}{cc}
A_{11^{\prime}}+\sum_2 A_{12}\left(A_{22}^{\text {diag }}\right)^{-1} A_{21^{\prime}}+\sum_2 B_{12}\left(A_{22}^{\text {diag }}\right)^{-1} B_{21^{\prime}} & A_{12} \\
A_{21} & A_{22}^{\text {diag }}
\end{array}\right), \\
\mathcal{B}_D^S=\left(\begin{array}{cc}
B_{11^{\prime}}+\sum_2 A_{12}\left(A_{22}^{\text {diag }}\right)^{-1} B_{21^{\prime}}+\sum_2 B_{12}\left(A_{22}^{\text {diag }}\right)^{-1} A_{21^{\prime}} & B_{12} \\
B_{21} & 0
\end{array}\right),
\end{gathered}
\end{equation}
ensuring the diagonal approximation of $A_{22}$ matrix at each step of the subtraction method. This approximation is particularly useful since it reduces the numerical complexity while retaining the dominant contributions of the \(2p\text{--}2h\) configurations.

\section{Results and discussion} \label{sec:results}

In the following, we present results for the isoscalar and isovector giant monopole (ISGMR and IVGMR) and quadrupole resonances (ISGQR and IVGQR) in $^{16}$O nucleus, calculated within several variants of the relativistic second random-phase approximation (RSRPA). The underlying single-particle basis is obtained from the relativistic Hartree equations, solved self-consistently in a spherical harmonic oscillator basis as described in Ref.~\cite{Nik}. For the GMR case, oscillator shells are included up to a principal quantum number \(N = 20\), while for the GQR calculations a reduced basis with \(N = 10\) is employed for both ground and excited states. Although this reduction is motivated primarily by computational constraints, it may contribute to the modest overestimation of the RSRPA excitation energies observed in the RSRPA results. 

The configuration space includes particle–hole (\(1p\text{--}1h\)) and two-particle--two-hole (\(2p\text{--}2h\)) excitations that satisfy the selection rules for total angular momentum, parity, and isospin. For the $1p\text{--}1h$  sector, unperturbed configurations with excitation energies up to 200~MeV are included in order to ensure completeness of the $1p\text{--}1h$ basis employed in calculations. In addition, transitions from occupied states to empty negative-energy Dirac states, denoted by \(\alpha\) and referred to as \(1\alpha\text{--}1h\) configurations, are incorporated with an extended energy cutoff of 2000~MeV. The inclusion of these \(1\alpha\text{--}1h\) contributions is essential to maintain consistency with the relativistic no-sea approximation and to eliminate unphysical spurious admixtures (see Ref. \cite{Niksic2002}) . For the \(2p\text{--}2h\) sector, an energy cutoff is introduced and systematically varied in order to verify the numerical convergence of strength distributions and the preservation of energy-weighted sum rules within the relativistic RSRPA framework. These truncation procedures are necessary to control the exponential increase of the configuration space dimensionality, while retaining the essential physical correlations relevant to the nuclear response.

In the present work the configuration space includes $1\alpha$--$1h$ configurations, while higher-order $2\alpha$--$2h$ configurations are not included explicitly. Although their inclusion would in principle extend the correlation space, exploratory calculations performed in a reduced model space indicate that these configurations do not lead to a systematic improvement in the description of the ISGMR strength distribution within the present subtraction scheme. For this reason, and in order to maintain computational feasibility, the present study is restricted to configurations including antiparticle states up to the $1\alpha$--$1h$ level.

\subsection{ISGMR}

\begin{figure}[ht]
\centering
\includegraphics[width=0.9\textwidth]{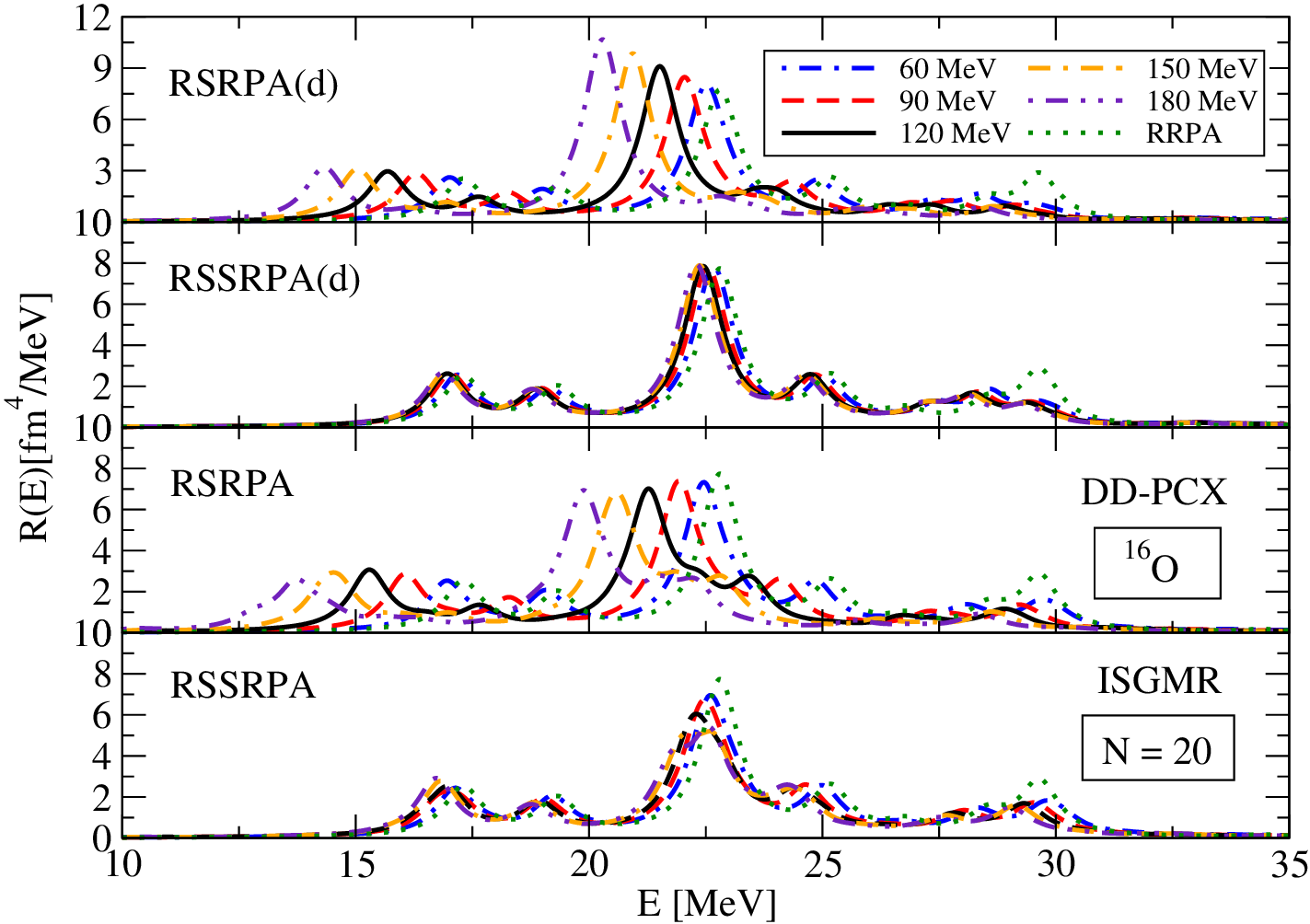}
\caption{Isoscalar monopole strength distributions for \textsuperscript{16}O calculated in the diagonal approximation and full RSRPA, without and with subtraction (see text for details).} 
\label{fig:ISGMR:N20}
\end{figure}

Figure~\ref{fig:ISGMR:N20} shows the calculated transition strength distribution of the isoscalar giant monopole resonance (ISGMR, \(0^+\)) for \textsuperscript{16}O, obtained using several variants of the RSRPA with density dependent point coupling interaction DD-PCX \cite{Yuk2019}. The ground state is described by solving the Dirac-Hartree equations in the basis constructed using spherical harmonic oscillator wave functions with a principal quantum number cutoff \(N = 20\). To facilitate a continuous spectral representation, the discrete transition strengths are folded with a Lorentzian of 1~MeV width.
The analysis includes both the standard RSRPA and subtracted version RSSRPA, considered in two implementations: one retaining the full residual interaction matrix, and another employing the diagonal approximation for the \(A_{22}\) block (Eq.~\ref{diagonal}), denoted with (d) in the legend of Fig.~\ref{fig:ISGMR:N20}. For comparison, the RRPA results are also displayed. In the fully self-consistent treatment, rearrangement terms associated with the density dependence of the interaction are incorporated in all submatrices (\(A_{11}\), \(A_{12}\), and \(A_{22}\)). In contrast, the diagonal approximation omits these contributions from the \(A_{22}\) sector.
The dependence of the results on the two-particle–two-hole configuration space is explored by varying the energy cutoff of \(2p\text{--}2h\) configurations across the values 60, 90, 120, 150, and 180~MeV.
As expected, enlarging the configuration space leads to a gradual downward shift in energy of the overall transition strength. When the subtraction procedure is applied, the resulting spectra from both RSSRPA and its diagonal counterpart, RSSRPA(d), exhibit numerical stability, with all prominent peaks of transition strength \(B(E_i) \gtrsim 0.01~\text{fm}^4\) converging between 150 and 180~MeV. The associated shifts in peak positions due to increasing the cutoff are small, typically \(\Delta E^{\mathrm{SSRPA}} \lesssim 150\)~keV and \(\Delta E^{\mathrm{SSRPA(d)}} \lesssim 120\)~keV.

These observations are consistent with previous non-relativistic studies~\cite{Yang2021,Gambacurta2015}, where the diagonal approximation was shown to introduce only small deviations when used in combination with the subtraction method. Our results confirm this behavior in the relativistic case, although with slightly more pronounced energy shifts, which are up to a few hundred keV, when comparing the fully self-consistent RSSRPA with RSSRPA(d) are observed in the fully self-consistent RSSRPA when compared to RSSRPA(d) for identical $2p\text{--}2h$ cutoffs.
We note that the definition of the diagonal approximation used in this work differs from that in Ref.~\cite{Gambacurta2015}. Here, the diagonal truncation is applied consistently in all stages of the calculation, including the matrix inversion required for the subtraction term. In contrast, Ref.~\cite{Gambacurta2015} employs the full \(A_{22}\) matrix except during the subtraction step, where the inversion is performed with only the diagonal part.

\begin{figure}[ht]
\centering
\includegraphics[width=0.9\textwidth]{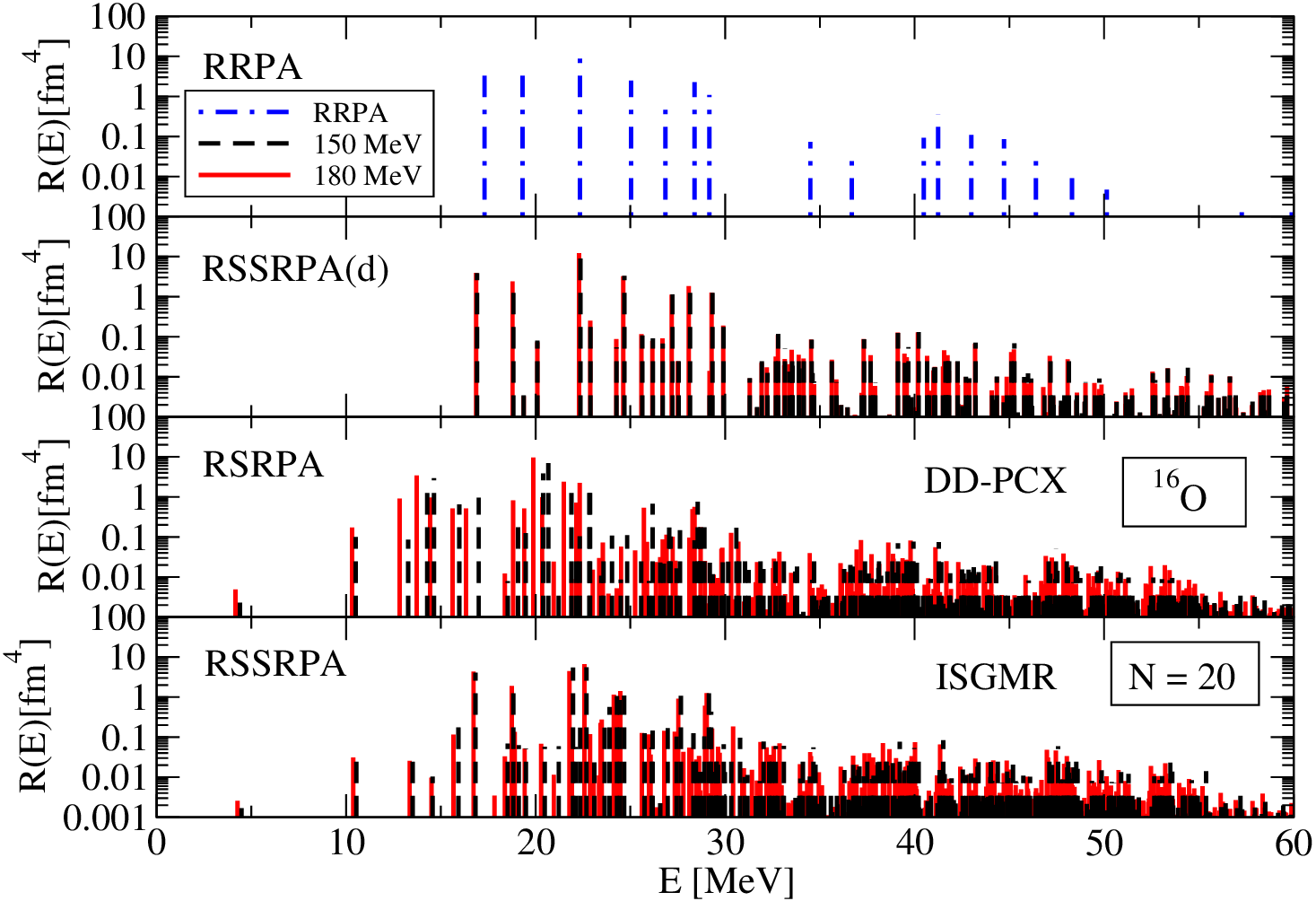}
\caption{Discrete isoscalar monopole transition strength distributions calculated with RRPA and several variants of RSRPA for \textsuperscript{16}O.}
\label{fig:ISGMR:N20:Discrete}
\end{figure}

Figure~\ref{fig:ISGMR:N20:Discrete} shows the discrete isoscalar monopole transition strength distributions for \textsuperscript{16}O, providing a more detailed view of the fragmentation pattern associated with the excitation spectrum. 
Compared to the simple RRPA, the inclusion of $2p$--$2h$ configurations in the RSRPA leads to a noticeable downward shift of dominant peaks in the transition strength distribution. For an energy cutoff of 150~MeV, the peaks with transition strengths \( B \gtrsim 0.1~\text{fm}^4 \) and predominantly $1p$--$1h$ character (\( \mathcal{N}_1 \gtrsim 0.8 \)) are lowered by up to 2~MeV. This shift becomes even more pronounced with a cutoff of 180~MeV, reaching nearly 2.5~MeV, which further differentiates the RSRPA spectrum from that of the RRPA. Few collective states emerge below 15 MeV, most of which exhibit a dominant two-particle–two-hole (\(2p\text{--}2h\)) character, quantified by normalized amplitudes \(\mathcal{N}_2 \gtrsim 0.99\) as defined in Eq.~\eqref{eq:srpa_norm_part}. However, the positions of these states remain stable as the 2p--2h energy cutoff increases. Only minor variations are observed in the transition strength, while the peak energies are essentially unaffected. 
Although the spectral redistribution also affects the intermediate-energy region, its impact there is more moderate. In this region, a few predominantly \(1p\text{--}1h\) states coexist with several mixed (\(0.2 \lesssim \mathcal{N}_2 \lesssim 0.8\)) and almost pure \(2p\text{--}2h\) states (\(\mathcal{N}_2 \gtrsim 0.99\)), while the latter become dominant type at higher excitation energies (\(E \gtrsim 40\)~MeV).
Even when the $2p$--$2h$ cutoff is extended to very high energies (beyond 150~MeV), the excitation energies in the standard RSRPA show no clear tendency toward convergence.

In contrast, the subtracted versions, RSSRPA and RSSRPA(d), display a more stable behavior, with the corresponding peak positions deviating by no more than approximately 1~MeV from the RRPA results. This deviation reflects the expected slight downward shift of the excitation energies, which tend to converge in both the full and diagonal formulations of RSSRPA for $2p$--$2h$ cutoff energies of \(E_{\text{cutoff}} \gtrsim 150~\text{MeV}\). The primary distinction between RSSRPA and RSSRPA(d) lies in the absence of residual interaction among \(2p\text{--}2h\) configurations in the latter. As a result, the fragmentation in RSSRPA(d) is somewhat less pronounced, leading to a reduced number of states with
predominantly \(2p\text{--}2h\) origin with low transition strengths (\(B \gtrsim 0.001~\text{fm}^4\)), distributed between the main collective peaks in the low and intermediate excitation energy region (\(E \lesssim 40\)~MeV).

Table~\ref{tab:gmr_2p2h} summarizes the number of $2p$--$2h$ configurations included in the ISGMR calculations as a function of the cutoff energy $E_{cutoff}$. The cutoff energies for the $1p$--$1h$ and $1\alpha$--$1h$ configuration spaces are kept fixed, corresponding to $N_{1p1h}=44$ and $N_{1\alpha1h}=64$. Consequently, the increase of the total configuration space is driven solely by the growing number of $2p$--$2h$ configurations. As expected, the number of configurations rises rapidly with increasing $E_{cutoff}$, which directly determines the dimension of the corresponding R(S)SRPA matrix. The eigenvalue problems, including the right eigenvectors, were solved using the LAPACK routine \texttt{DGEEV} for real non-symmetric matrices \cite{lapack99, lapack_web}, with a computational cost that scales approximately as $\mathcal{O}(N_{dim}^{3})$ with the matrix dimension ($N_{dim}$). In practice, reasonable convergence of the calculated observables is already achieved for $2p$--$2h$ cutoff energies in the range of 150--180 MeV.

\setlength{\tabcolsep}{6pt}
\begin{table}[ht]
\centering
\caption{Number of $2p$--$2h$ configurations included in the GMR calculations  as a function of the $2p$--$2h$ cutoff energy $E_{cutoff}$. The cutoff energies for the $1p$--$1h$ and $1\alpha$--$1h$ configuration spaces are kept fixed, corresponding to $N_{1p1h}+N_{1\alpha1h}=108$. The table also lists the resulting total number of configurations $N_{\mathrm{tot}}$ and the corresponding dimension of the R(S)SRPA matrix $N_{\mathrm{dim}}=2N_{\mathrm{tot}}$.} 
\begin{tabular}{c|c c c c c c c c c}
\hline\hline
$E_{cutoff}$ [MeV] & 30.0 & 60.0 & 90.0 & 120.0 & 150.0 & 180.0 & 210.0 & 240.0 & 270.0 \\
\hline
$N_{2p2h}$ & 9 & 408 & 1890 & 4326 & 7390 & 10797 & 14691 & 18611 & 22700 \\
$N_{\mathrm{tot}}$ & 117 & 516 & 1998 & 4434 & 7498 & 10905 & 14799 & 18719 & 22808 \\
$N_{\mathrm{dim}}$ & 234 & 1032 & 3996 & 8868 & 14996 & 21810 & 29598 & 37438 & 45616 \\
\hline\hline
\end{tabular}
\label{tab:gmr_2p2h}
\end{table}

The energy-weighted moments $m_0$, $m_1$, and $m_2$ (see eq. \ref{moments}), which characterize different aspects of the transition strength distributions shown in Fig.~\ref{fig:ISGMR:N20:Discrete} are given in Tab.~\ref{tabIS0}. The zeroth moment $m_0$ quantifies the total transition strength, regardless of the excitation energy. When comparing the full energy interval, the differences between RRPA and RSSRPA(d) or RSSRPA in $m_0$ are negligible (less than 0.15\%). However, within the experimental energy window 11-40 MeV \cite{Lui2001}, RSSRPA(d) shows a 3.6\% decrease relative to RRPA, and RSSRPA yields an even larger reduction of 5.4\%. For the energy-weighted moment $m_1$, both extended models exhibit a moderate increase compared to RRPA, i.e., RSSRPA(d) (RSSRPA) exceeds the RRPA value by $8.8\%$ ($10.4\%$). In Ref.~\cite{Yang2021}, the total energy-weighted sum \( m_1 \) was found to increase by approximately 7.5\% (9.5\%) in the SSRPA(d) (SSRPA) calculation with Skyrme interaction compared to the corresponding RPA result, showing a trend consistent with our results.
The second moment $m_2$ reveals the most pronounced difference, with RSSRPA reaching almost double the RRPA value in the full spectrum. Nevertheless, in the experimental window, $m_2$ for both RSSRPA(d) and RSSRPA is reduced compared to RRPA. These trends indicate a redistribution of the strength to higher excitation energies beyond the existing experimental range, and enhanced fragmentation in these models.

The inverse energy-weighted moment $m_{-1}$ is sensitive to the low-energy part of the spectrum and plays an essential role in sum-rule evaluations, especially in the context of compressional modes. Despite the inclusion of $2p$--$2h$ configurations, the value of $m_{-1}$ remains remarkably stable across all models. Variations are below 0.1\%, indicating $m_{-1}$ is rather robust quantity in theoretical considerations (see Tab. \ref{tabIS0}).
Furthermore, it indicates that all approaches reproduce the same static isoscalar polarizability, i.e., the same nuclear incompressibility. 
Such stability of $m_{-1}$ implies that the theoretical frameworks are mutually self-consistent, and that the differences between models affect only the dynamical redistribution of the strength rather than the total magnitude of the static monopole response. 
A similar behavior has been observed in nonrelativistic frameworks, where the $m_{-1}$ moment remains unchanged between the SSRPA and RPA calculations~\cite{Yang2021}. However, a noticeable deviation arises in non-relativistic models when the diagonal approximation is employed, indicating a partial loss of self-consistency that originates from the omission of interactions among \(2p\text{--}2h\) configurations in the inverse calculation of the \(A_{22}\) matrix during subtraction procedure, i.e., calculation of $A_{11}^S$, while keeping $A_{22}$ in the full form in $\mathcal{A}_D^S$ \cite{Gambacurta2015, Yang2021}. On the other hand, in our calculations, the diagonal approximation of $A_{22}$ was employed throughout the entire subtraction procedure.
However, the constrained energy, defined as $\sqrt{m_1/m_{-1}}$, increases from 23.856 MeV in RRPA to 25.058 MeV in RSSRPA. Within the experimental window \cite{Lui2001}, this ratio drops to values closer to the experimental result of 19.63 MeV, with the smallest deviation to the experiment in the RSSRPA case. This behavior confirms the influence of higher-energy components and underscores the importance of subtraction for regulating effective interactions.
\begin{table}[ht]
\begin{center}
    \caption{Energy-weighted moments, derived excitation energies and energy weighted sum rule (EWSR) of isoscalar monopole transitions
    obtained by the RRPA, RSSRPA(d) and RSSRPA calculations for $^{16}$O with DD-PCX interaction. For RSSRPA(d) and RSSRPA calculations the energy $2p$--$2h$ cutoff is taken $E_{cutoff}=150$ MeV. The values in round brackets are calculated within experimental window from \cite{Lui2001}. } 
    \begin{tabular}{c c c c c}
        \hline\hline
         $ $ & RRPA & RSSRPA(d) & RSSRPA & Exp. \cite{Lui2001}\\
        \hline
        $m_0$ [fm$^4$] & 29.794 (28.780) & 29.789 (27.736) &  29.838 (27.235) &   \\
        $m_1$ [MeVfm$^4$]  & 726.357 (680.288) &  790.074 (638.375) & 801.735 (624.961)  &  \\
        $m_2$ [MeV$^2$fm$^4$] & 18725.64 (16545.340) & 28768.79 (15175.80) & 29548.40 (14897.78) & \\
        $m_{-1}$ [MeV$^{-1}$fm$^4$] & 1.276 (1.254) & 1.276 (1.243) & 1.277 (1.230) & \\
        $m_1/m_0$[MeV] & 24.380 (23.638) & 26.522 (23.016) & 26.870 (22.947) & 21.13$\pm$0.49\\ 
        $\sqrt{m_1/m_{-1}}$ [MeV] & 23.856 (23.296) & 24.879 (22.661) & 25.058 (22.539)  & 19.63$\pm$0.38\\
        $\sqrt{m_3/m_{1}}$ [MeV] & 26.873 (24.659) & 47.860 (24.178) & 48.011 (24.329)  & 24.89$\pm$0.59\\
        $\Delta$ [MeV] & 5.843 (4.020) & 16.196 (4.172) & 16.380 (4.522) & 8.76$\pm$1.82 \\
        EWSR[\%] & 99.67 (93.35) & 108.41 (87.60) & 110.01 (85.76) & 48$\pm$10 \\
        \hline\hline
    \end{tabular}
    \label{tabIS0}
\end{center}
\end{table}

The centroid energy  $m_1/m_0$ provides a characteristic measure of the average excitation energy. The RRPA value of 24.380 MeV increases to 26.522 MeV in RSSRPA(d) and 26.870 MeV in RSSRPA, representing a relative increase of about 10\%. The increase in the ratio \( m_1/m_0 \) when the full positive spectrum is considered can be attributed to the presence of a larger number of mixed and predominantly \(2p\text{--}2h\) states at excitation energies above 40~MeV. These states contribute with small but finite transition strengths, leading to an overall enhancement of the \(m_1\) moment relative to the RRPA result, while the \(m_0\) moment remains practically unchanged.
Experimentally, the centroid is observed at 21.13 MeV, which is substantially lower than the theoretical predictions. However, when the analysis is performed within the experimental energy range, the calculated $m_1/m_0$ values reach closer to the experimental centroid energy.
The ratio $\sqrt{m_3/m_1}$, sensitive to high-energy tails in the strength function, shows a drastic increase in the extended models, from 26.873 MeV in RRPA to nearly 48 MeV in RSSRPA. This shift reflects additional strength obtained at high excitation energies, a typical consequence of including complex $2p$--$2h$ correlations. Nevertheless, within the experimental window \cite{Lui2001}, all theoretical values converge toward the measured value of 24.89 MeV (see Tab.~\ref{tabIS0}), suggesting acceptable agreement when limited to observed excitations.

The energy-weighted sum rule (EWSR) provides a global consistency check of the model \cite{Hartmann2002}. In RRPA, it is nearly saturated with 99.67\% of the expected strength. RSSRPA(d) and RSSRPA exceed this limit, yielding 108.41\% and 110.01\%, respectively. The obtained enhancement in the EWSR follows the same trend as $m_1$, further supporting the consistency of our results with those reported in Ref.~\cite{Yang2021}. Within the experimental window \cite{Lui2001}, the EWSR values are significantly reduced: RRPA yields 93.35\%, RSSRPA(d) 87.60\%, and RSSRPA 85.76\%. These values should be compared with the experimental estimate of 
$(48 \pm 10)\%$ of the EWSR \cite{Lui2001}. The discrepancy implies that a considerable portion of the strength lies outside the measured domain or is experimentally inaccessible. Subtracted approaches provide improved consistency with sum rules while retaining enhanced collectivity. However, even within the RSSRPA framework, the calculated exhaustion of the ISGMR EWSR exceeds the experimental value by more than 20\%, suggesting that correlations beyond the two-particle--two-hole ($2p$--$2h$) space, such as three-particle--three-hole ($3p$--$3h$) configurations, may be essential for a more accurate description of the experimental strength distribution.
Although, $3p$--$3h$ configurations could potentially reduce EWSR by few percent, certain correlation effects on the ground-state level, which are absent in the RH approximation, can induce noticeable shifts in excitation energies and influence the fragmentation of transition strength \cite{PhysRevC.87.054330}.
However, theoretical studies indicate that the direct impact of $3p$--$3h$ configurations on the total EWSR is likely limited to a few percent \cite{Ponomarev1996, Colo2022_methods}. Instead, a more significant effect arises from the treatment of the ground state, i.e., the degree of correlations already present in the reference state. As discussed in~\cite{PhysRevC.87.054330}, neglecting such correlations in the RH approximation can lead to an overestimation of excitation energies, EWSR and a partially inaccurate fragmentation pattern. 

\subsection{IVGMR}

\begin{figure}[ht]
\centering
\includegraphics[width=0.9\textwidth]{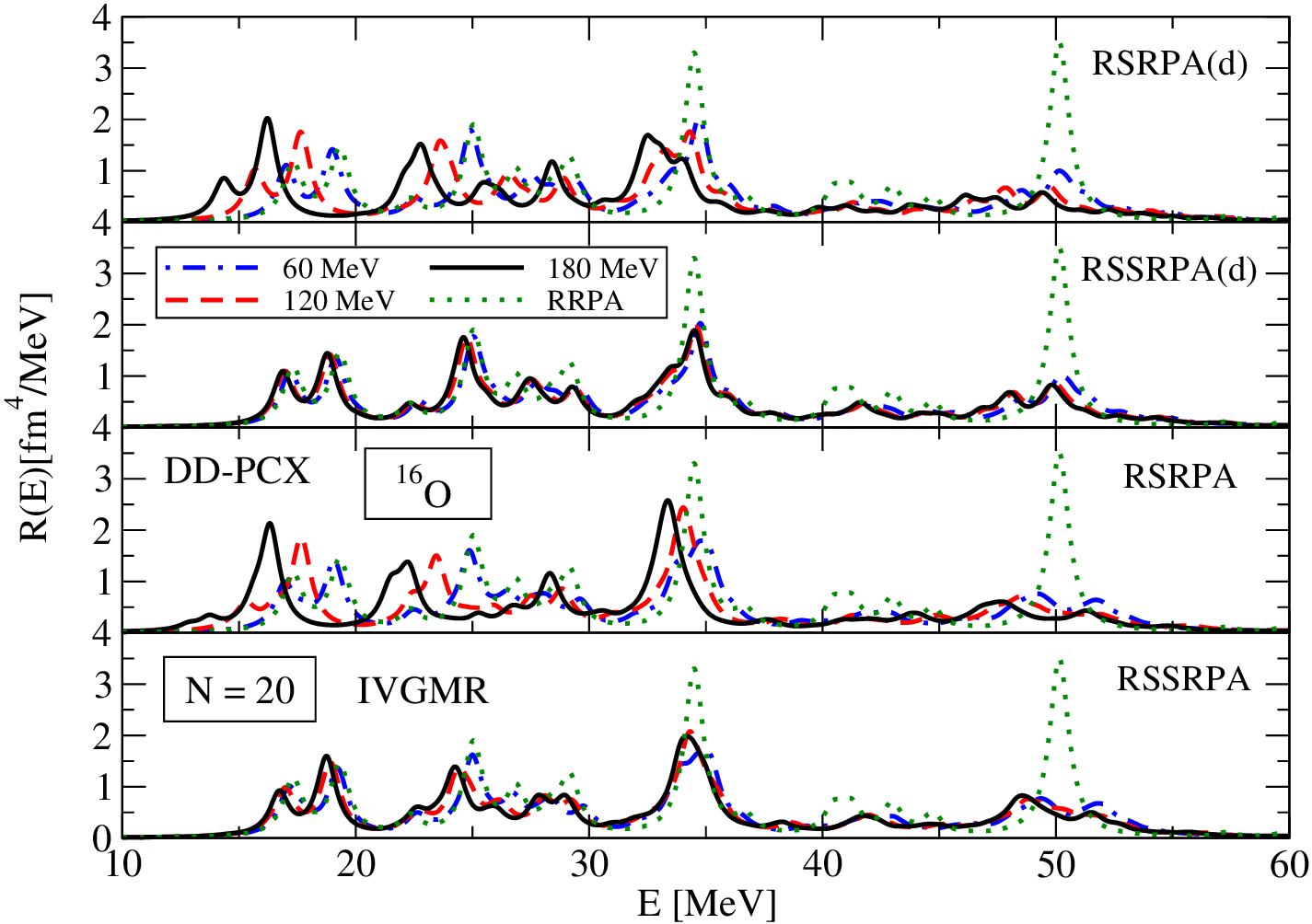}
\caption{Isovector monopole strength distributions calculated by RRPA and several variants of RSRPA for \textsuperscript{16}O, shown for several $2p$--$2h$ cutoffs.}
\label{fig:IVGMR:N20}
\end{figure}

\begin{figure}[ht]
\centering
\includegraphics[width=0.9\textwidth]{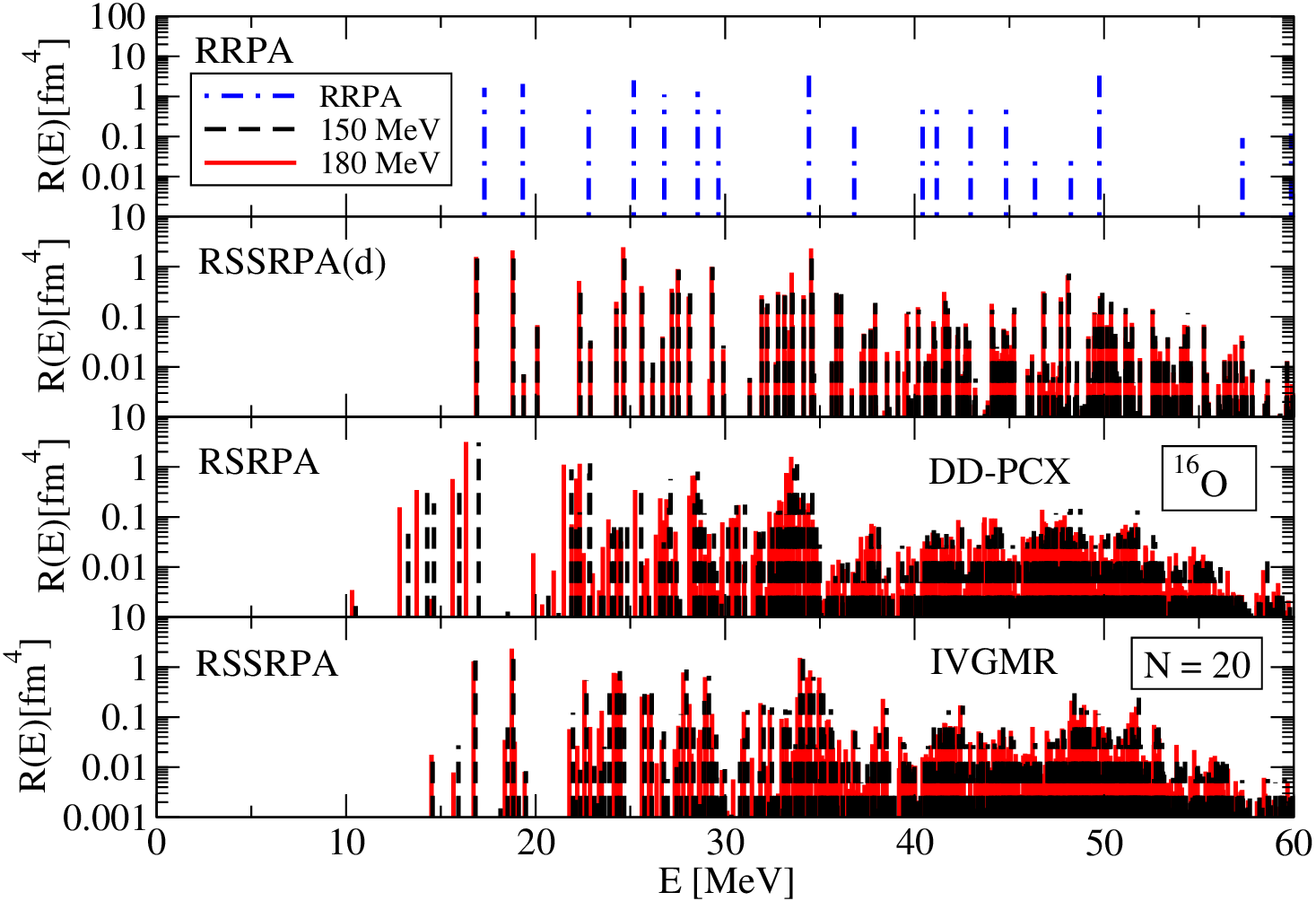}
\caption{Discrete isovector monopole strength distributions calculated by RRPA and several variants of RSRPA for \textsuperscript{16}O.}
\label{fig:IVGMR:N20:Discrete}
\end{figure}

In this section, we analyze the properties of IVGMR in \(^{16}\)O, again by using several variants of RSRPA based on the DD-PCX interaction. The transition strength distributions obtained using the RRPA, RSRPA(d), RSSRPA(d), RSRPA, and RSSRPA are shown in Fig.~\ref{fig:IVGMR:N20} for different values of the \(2p\text{--}2h\) energy cutoff, specifically 60, 120, and 180 MeV. In order to illustrate the fine structure of the excitations, the corresponding discrete spectra for selected cutoffs (150 and 180 MeV) are displayed in Fig.~\ref{fig:IVGMR:N20:Discrete}. The RRPA  yields prominent peaks located at higher excitation energies ($E\gtrsim30$ MeV).
Upon inclusion of \(2p\text{--}2h\) configurations via the RSRPA approach, the overall fragmentation pattern in Fig.~\ref{fig:IVGMR:N20} remains largely unchanged, although a noticeable shift of peaks below 30~MeV towards lower energies is observed with increasing \(2p\text{--}2h\) cutoff energy, accompanied by moderate variations in strength. 
However, the energy region above \(E \gtrsim 30\)~MeV reveals a redistribution of strength, where several sharp peaks are diminished and replaced by weaker transitions, for all RSRPA variants, including the subtracted and non-subtracted forms, in both the full and diagonal implementations. 
On the other hand, a closer examination of the RSRPA discrete strength distributions reveals, in comparison with the RRPA spectrum (Fig.~\ref{fig:IVGMR:N20:Discrete}), an increased number of mixed and predominantly \(2p\text{--}2h\) states that appear together with the \(1p\text{--}1h\)-dominated states, particularly above \(E \gtrsim 20\)~MeV. The number of mixed and predominantly \(2p\text{--}2h\) states with transition strengths above \(10^{-3}\,\mathrm{fm}^4\) does not change significantly as the \(2p\text{--}2h\) cutoff energy increases from 150~MeV to 180~MeV. 
However, even at such high cutoff values, convergence of the RSRPA for the IVGMR is not achieved, as shifts of up to nearly 0.5~MeV in the positions of several states remain present, particularly in the low-energy region (below 20~MeV) and, to a lesser extent, in the intermediate part of the spectrum (between 20 and 40~MeV).

This fragmentation reflects the enhanced configuration mixing and spreading effects induced by complex correlations. When the subtraction method is applied, as in RSSRPA and RSSRPA(d), the transition strength becomes more stabilized, with less pronounced fragmentation and more localized peaks.
In particular, the RSSRPA(d) calculation, which omits residual interactions between \(2p\text{--}2h\) states, results in reduced complexity of the excitation spectrum and transition strength is more concentrated in predominantly \(1p\text{--}1h\) states ($\mathcal{N}_1 \gtrsim 0.8$). At a cutoff energy of 150 MeV, RSRPA peak positions differ from RRPA by up to approximately 2 MeV, while the corresponding shifts in RSSRPA and RSSRPA(d) remain below 1 MeV (see Fig.~\ref{fig:IVGMR:N20:Discrete}).
As the cutoff increases to 180 MeV, the peak positions in RSRPA continue to move downwards by about 1 MeV, whereas RSSRPA and RSSRPA(d) show only minor additional shifts, $\approx$100 keV in both cases, indicating convergence of the excitation spectrum in the subtracted formulations.

Both ISGMR and ISGQR are more easily identified experimentally because they exhibit strong collective behavior, producing well-defined and concentrated strength distributions that dominate the inelastic scattering spectra at small angles, particularly in \((\alpha,\alpha')\) reactions, where isoscalar transitions of low multipolarity are selectively excited \cite{Lui2001}. Their energy centroids are well separated and the corresponding multipole decomposition analysis allows for a clear isolation of these modes. In contrast, the isovector giant monopole resonance (IVGMR) is significantly weaker, i.e., typically suppressed by at least one order of magnitude, which severely limits its visibility in experimental spectra. As a consequence, dedicated measurements of the IVGMR are scarce, and systematic experimental data for this mode remain largely unavailable.

\subsection{ISGQR}

\begin{figure}[ht]
\centering
\includegraphics[width=0.9\textwidth]{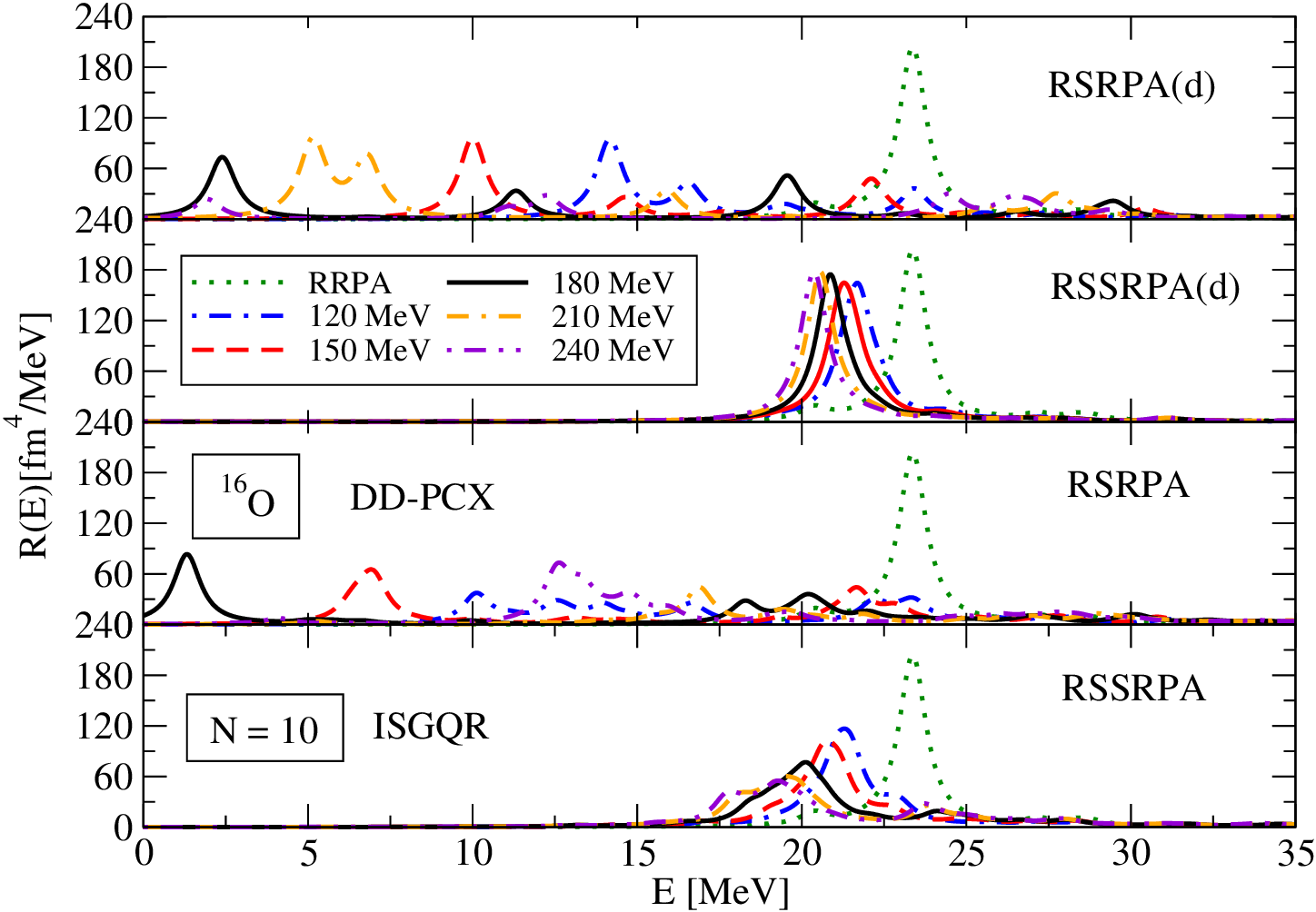}
\caption{The ISGQR strength distribution for \textsuperscript{16}O, calculated with the RRPA and several variants of the RSRPA, for the range of $2p$--$2h$ cutoffs given in the legend.}
\label{fig:ISGQR}
\end{figure}

\begin{figure}[ht]
\centering
\includegraphics[width=0.9\textwidth]{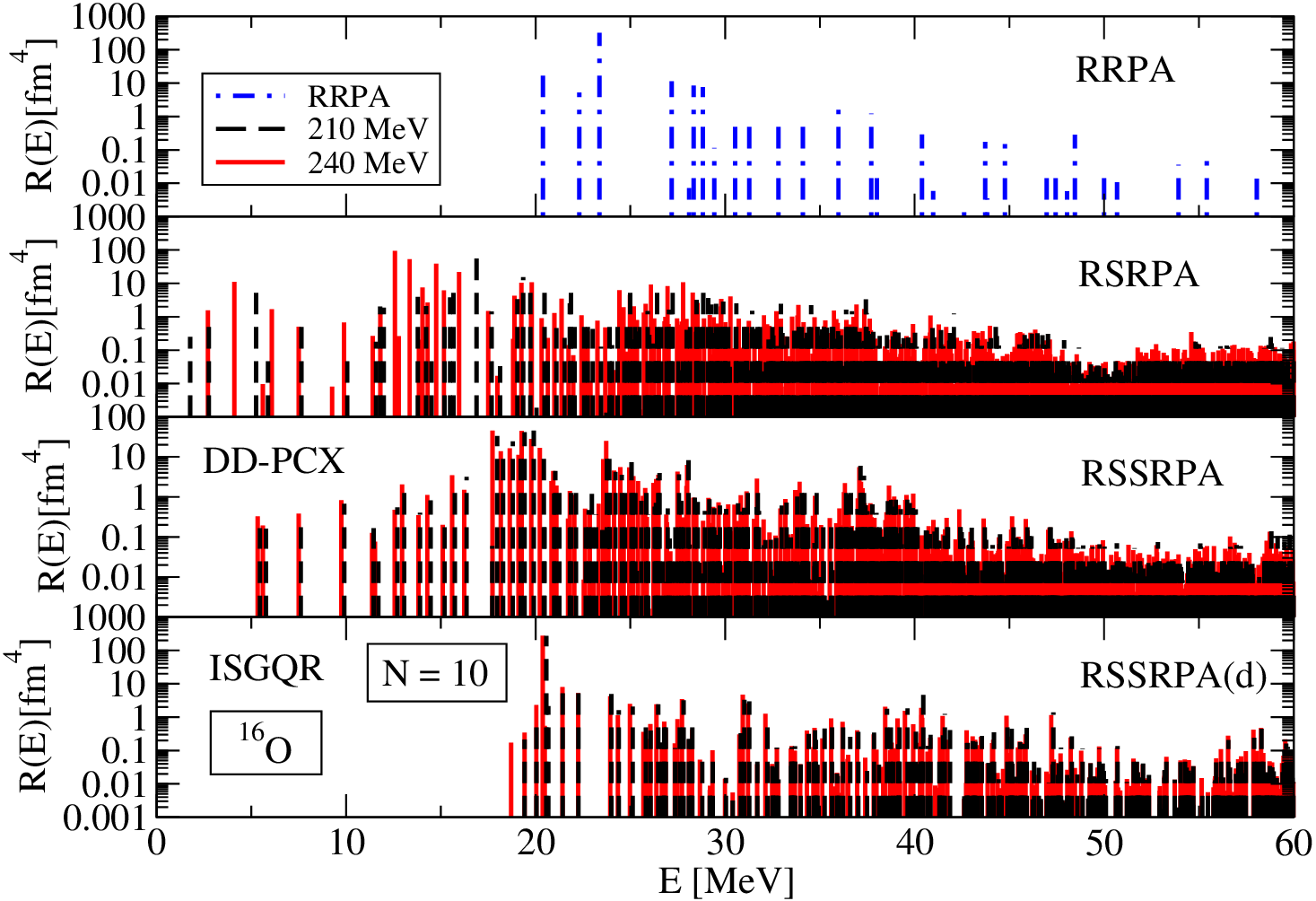}
\caption{Discrete ISGQR strength distribution for $^{16}$O, calculated with RRPA and several variants of RSRPA, using two choices of $2p$--$2h$ cutoffs.}
\label{fig:ISGQR:Discrete}
\end{figure}

Figure~\ref{fig:ISGQR} displays the ISGQR strength distributions in $^{16}$O as obtained using the RRPA, RSRPA, RSSRPA, and RSSRPA(d) with energy cutoffs for 2p--2h configurations ranging from 120 to 240 MeV. Discrete transition strengths for cutoff values of 210 and 240 MeV are shown separately in Fig.~\ref{fig:ISGQR:Discrete}. These calculations are performed using a reduced harmonic oscillator basis limited to a principal quantum number $N = 10$, due to computational constraints. The inclusion of $2p$--$2h$ correlations through the RSRPA formalism results in a pronounced fragmentation of the strength distribution. Already for low cutoff values, such as 60 MeV, the dominant peak observed in RRPA splits into several components in RSRPA, accompanied by a noticeable shift of strength to lower energies.

In comparision to the monopole modes (ISGMR and IVGMR), the ISGQR transition spectrum exhibits enhanced sensitivity to the couplings between $1p$--$1h$ and $2p$--$2h$ configurations, manifested in both stronger fragmentation and more pronounced downward energy shifts in RSRPA relative to RRPA. Unlike the case of ISGMR, where the diagonal approximation RSRPA(d) closely reproduces the full RSRPA results, for ISGQR this approximation is less reliable. For large 2p--2h energy cutoffs (typically \( E_{\text{cutoff}} \gtrsim 210~\text{MeV} \)), the RSRPA calculations may yield complex eigenvalues for the lowest few excitation modes. Such behavior was reported in non-relativistic SRPA calculations in Ref. \cite{Papakonstantinou2010}, where authors observed up to four complex eigenvalues in full SRPA calculations using realistic interactions. This violation of the stability condition arises from the lack of ground-state correlations and the non-positive definiteness of the SRPA matrix. A similar mechanism likely underlies the emergence of complex-conjugate pair observed in our RSRPA calculations for $2p$--$2h$ energy cutoffs $E_{cutoff}\gtrsim210$ MeV.

The implementation of the subtraction procedure in RSSRPA and RSSRPA(d) leads to more stable spectra. It removes the spurious low-energy behavior and ensures a better convergence pattern. In RSSRPA, the dominant transition peak becomes fragmented into multiple states located between roughly 15 and 25 MeV. In addition, there are weak transitions shaped by a small set of nearly pure $2p$--$2h$ states ($\mathcal{N}_2 \gtrsim 0.99$), which cluster below $E_x \lesssim 17$ MeV and remain relatively insensitive to variations in the $2p$--$2h$ cutoff. For cutoff values $E_\text{cutoff} \lesssim 150$ MeV, the positions of the excitation energies of the predominantly $1p$--$1h$ and mixed states decrease by a few hundred keV for every 30 MeV increase in the cutoff. Beyond 180 MeV, the energy shift stabilizes, and for $E_{cutoff} \gtrsim 200$ MeV energy shift toward lower energies is $\Delta E_x \lesssim 50$ keV per 10 MeV, indicating reasonable convergence of RSSRPA. 
This trend is consistent with findings in nonrelativistic SSRPA calculations using Skyrme energy density functionals~\cite{Gambacurta2015}.  In contrast, RSSRPA(d) results are more concentrated, with the majority of the strength carried by a single dominant state. Although this indicates less fragmentation, the convergence of the peak position is similarly efficient. For example, in the 90--240 MeV cutoff range, the dominant peak shifts downward by about 30\% every 30 MeV. 
This behavior underscores the importance of treating $2p$--$2h$ correlations and their mutual interaction in the full form in order to achieve a more accurate description of collective excitations.

For the ISGQR channel, the cutoff energies for the $1p$--$1h$ and $1\alpha$--$1h$ configuration spaces are kept fixed, corresponding to $N_{1p1h}=68$ and $N_{1\alpha1h}=82$, respectively. 
The scaling of the configuration space with increasing $2p$--$2h$ cutoff energy is illustrated in Table~\ref{tab:gqr_2p2h}. 
For illustration purposes, the number of $2p$--$2h$ configurations is shown up to $E_{cutoff}=300$ MeV, where the corresponding R(S)SRPA matrix dimensions approach $\sim 5\times10^{4}$. 
However, the full calculations presented in this work were performed up to $E_{cutoff}=240$ MeV, where a reasonable convergence of the results was achieved.

\begin{table}[ht]
\centering
\caption{Number of $2p$--$2h$ configurations included in the GQR calculations as a function of the $2p$--$2h$ cutoff energy $E_{cutoff}$. The cutoff energies for the $1p$--$1h$ and $1\alpha$--$1h$ configuration spaces are kept fixed, corresponding to $N_{1p1h}+N_{1\alpha1h}=150$. The table also lists the resulting total number of configurations $N_{\mathrm{tot}}$ and the corresponding dimension of the R(S)SRPA matrix $N_{\mathrm{dim}}=2N_{\mathrm{tot}}$.}
\begin{tabular}{c|cccccccccc}
\hline\hline
$E_{cutoff}$ [MeV] & 30.0 & 60.0 & 90.0 & 120.0 & 150.0 & 180.0 & 210.0 & 240.0 & 270.0 & 300.0 \\
\hline
$N_{2p2h}$ & 24 & 505 & 2520 & 5980 & 9841 & 14366 & 19062 & 22022 & 23607 & 24353 \\
$N_{\mathrm{tot}}$ & 174 & 655 & 2670 & 6130 & 9991 & 14516 & 19212 & 22172 & 23757 & 24503 \\
$N_{\mathrm{dim}}$ & 348 & 1310 & 5340 & 12260 & 19982 & 29032 & 38424 & 44344 & 47514 & 49006 \\
\hline\hline
\end{tabular}
\label{tab:gqr_2p2h}
\end{table}

The zeroth and first energy moments, $m_0$ and $m_1$, calculated up to the eigenvalue $E_x < 150$ MeV, are shown in Tab.~\ref{tabIS2}. The $m_0$ moment decreases slightly from 382.590 fm$^4$ in RRPA to 364.592 fm$^4$ in RSSRPA(d) and 369.347 fm$^4$ in RSSRPA, i.e., by roughly 4.7\% and 3.5\%. Within the experimental window (11--40 MeV), the reductions are more substantial: RSSRPA(d) shows an absolute drop of $\approx$49 fm$^4$ and RSSRPA about 42 fm$^4$, which correspond to relative reductions of 12.9\% and 11.0\%, respectively. Similarly, non-relativistic SSRPA calculations employing the Skyrme interaction show relative increase of about 15\% in the \( m_1 \) moment compared to RPA as the $2p$--$2h$ energy cutoff is extended up to 50~MeV, while the \(m_0\) moment remains nearly unchanged, exceeding the RPA value by only about 0.5\%.~\cite{Gambacurta2015}. More recent SSRPA calculations with Skyrme interactions including tensor terms report an enhancement of approximately 17.4\% in the diagonal approximation and 7.4\% in the full calculation~\cite{Yang2021}, following the same general trend observed in the present work.
For the energy-weighted moment $m_1$, the full-spectrum values are similar across all models, although RSSRPA shows a 7.9\% increase relative to RRPA. However, within the experimental window, RSSRPA(d) and RSSRPA exhibit reductions of around 20.4\% and 14.1\%, respectively, indicating that some transition strength is shifted outside the observable energy domain.

\begin{table}[ht]
\begin{center}
    \caption{Energy-weighted moments, derived excitation energies and EWSR of isoscalar $2^+$  transitions obtained with RRPA, RSSRPA(d) and RSSRPA for $^{16}$O using DD-PCX interaction. For RSSRPA(d) and RSSRPA calculations the energy $2p$--$2h$ cutoff is taken $E_{cutoff}=210$ MeV. The values in round brackets and $\Delta$ are calculated within experimental window from \cite{Lui2001}.}
    \begin{tabular}{c c c c c}
        \hline\hline
         $ $ & RRPA & RSSRPA(d) & RSSRPA & Exp. \cite{Lui2001}\\
        \hline
        $m_0$ [fm$^4$] & 382.590 (380.987) &  364.592 (331.881) &  369.347 (338.829)  &   \\
        $m_1$ [MeVfm$^4$]  & 9151.344 (9057.081) & 9356.759 (7205.863) & 9882.231 (7782.525) &  \\
        $m_{-1}$ [MeV$^{-1}$fm$^4$] & 16.167 (16.137) & 16.125 (15.546) & 16.230 (15.591) & \\
        $m_1/m_0$ [MeV] & 23.919 (23.773) & 25.664 (21.712) & 26.756 (22.969) & 19.76$\pm$0.22\\ 
        $\sqrt{m_1/m_{-1}}$ [MeV] & 23.792 (23.691) & 24.089 (21.529) & 24.675 (22.342)  & \\
        $\sqrt{m_3/m_{1}}$ [MeV] & 25.222 (24.112)& 45.672 (22.687) & 46.927 (25.479) & \\
        $\Delta$ [MeV] & 2.216 & 3.483 & 6.026 & 5.11$\pm$ 0.17 \\
        EWSR[\%] & 100.22 (99.19) & 102.47 (78.82) & 108.22 (85.23) & 53$\pm$10 \\
        \hline\hline
    \end{tabular}
    \label{tabIS2}
\end{center}
\end{table}

The inverse moment $m_{-1}$ remains nearly unchanged across models, varying by less than 0.3\% in the full spectrum and less than 4\% in the experimental window. This consistency indicates that the low-energy response is not significantly affected by the inclusion of $2p$--$2h$ configurations. 
For the ISGQR, a similar stability of $m_{-1}$ demonstrates that all approaches yield nearly identical static quadrupole polarizabilities, implying that model-dependent effects modify only the detailed strength distribution rather than the overall quadrupole response. 
This confirms that the underlying physics of nuclear elasticity in the quadrupole transitions is preserved across different theoretical formulations.
In contrast, the constrained energy $\sqrt{m_1/m_{-1}}$ increases from 23.792 MeV in RRPA to 24.089 MeV in RSSRPA(d) and 24.675 MeV in RSSRPA. Within the experimental window, it decreases closer to the experimental value of 19.63 MeV \cite{Lui2001}, suggesting that the models still overestimate the stiffness of the response, but improvements are present in the truncated spectrum (see Tab.~\ref{tabIS2}).

The centroid energy $m_1/m_0$ shows a systematic increase in extended models, from 23.919 MeV in RRPA toward 26.756 MeV in RSSRPA. In the experimental window, the RSSRPA value shifts downward to 22.969 MeV, approaching the experimental value of 19.76 MeV, thus improving the RRPA result. 
A similar trend is reported in Ref.~\cite{Gambacurta2015}, where, within the nonrelativistic Skyrme interaction, the \(m_1/m_0\) moment increases by approximately 10\% (15\%) for the \(2p\text{--}2h\) energy cutoff \(E_{\text{cutoff}} = 40\)~(50)~MeV when compared with RPA value.
The high-order ratio $\sqrt{m_3/m_1}$ increases considerably in the full spectrum for RSSRPA(d) and RSSRPA, reaching values of 45.672 MeV and 46.927 MeV, i.e., nearly twice the RRPA value. These values indicate significant redistribution of strength to higher energies. 

The width $\Delta$, which characterizes the fragmentation of the transition strength, increases consistently with model complexity. In RRPA, the width is only 2.216 MeV, while in RSSRPA(d) and RSSRPA it grows to 3.483 MeV and 6.026 MeV, respectively. The broader distribution in RSSRPA is consistent with the expected spreading due to complex configurations. Experimental measurements suggest a width of 5.11 MeV \cite{Lui2001}, which is most closely matched by the RSSRPA model. This suggests that including $2p$--$2h$ configurations not only improves the energy centroids but also leads to more realistic fragmentation patterns.

The energy-weighted sum rule (EWSR) for the isoscalar giant quadrupole resonance (ISGQR) is nearly fully exhausted in all three models, yielding 100.22\%, 102.47\%, and 108.22\% for RRPA, RSSRPA(d), and RSSRPA, respectively, when evaluated over the full excitation spectrum. 
Similar findings have been reported in Ref.~\cite{Gambacurta2015} for the Skyrme-based SSRPA approach, where the first energy moment increases to approximately \( m_1^{\text{SSRPA}} \approx 1.15\,m_1^{\text{RPA}} \), reflecting an enhanced exhaustion of the energy-weighted sum rule (EWSR), even at a relatively low \(2p\)--\(2h\) energy cutoff of \(E_{\text{cutoff}} = 50\)~MeV. This enhancement can be attributed to the presence of a large number of \(2p\)--\(2h\) configurations (\(\mathcal{N}_2 \gtrsim 0.99\)), which, despite being associated with very small transition strengths (\(B \lesssim 0.001\) fm$^{4}$), still contribute to the \(m_1\) moment due to their energy weighting, thus directly influencing the total EWSR.
However, when the integration is restricted to the experimental window of $11$–$40$~MeV \cite{Lui2001}, the EWSR values decrease by approximately $1\%$ in RRPA, and by as much as $24\%$ and $23\%$ in RSSRPA(d) and RSSRPA, respectively, indicating that a substantial fraction of the strength is redistributed to energy regions outside the experimental range due to the coupling with $2p$--$2h$ configurations.

\subsection{IVGQR}
\begin{figure}[ht]
\centering
\includegraphics[width=0.9\textwidth]{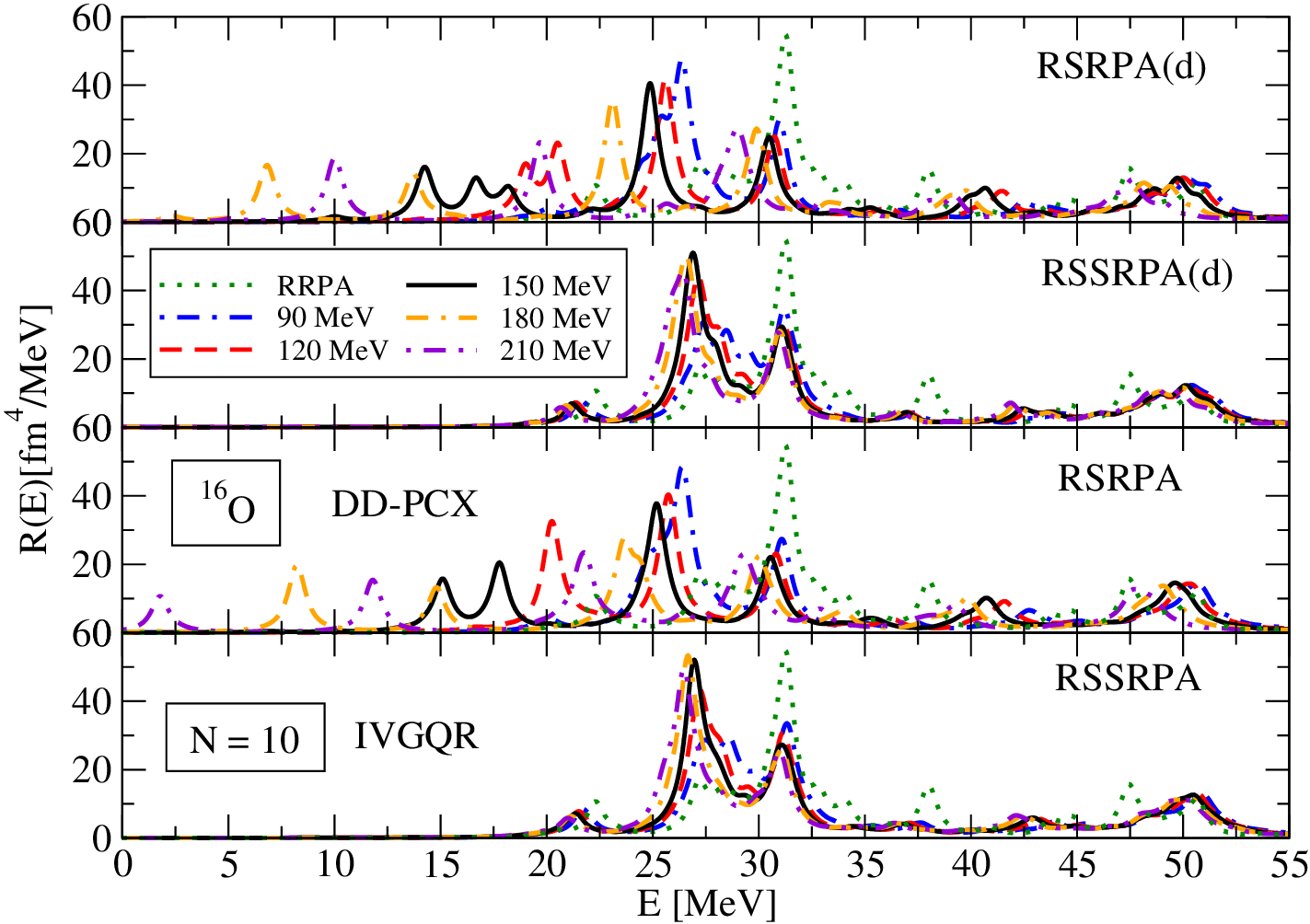}
\caption{IVGQR strength distribution calculated in the RSRPA framework for \textsuperscript{16}O.}
\label{fig:IVGQR}
\end{figure}

\begin{figure}[ht]
\centering
\includegraphics[width=0.9\textwidth]{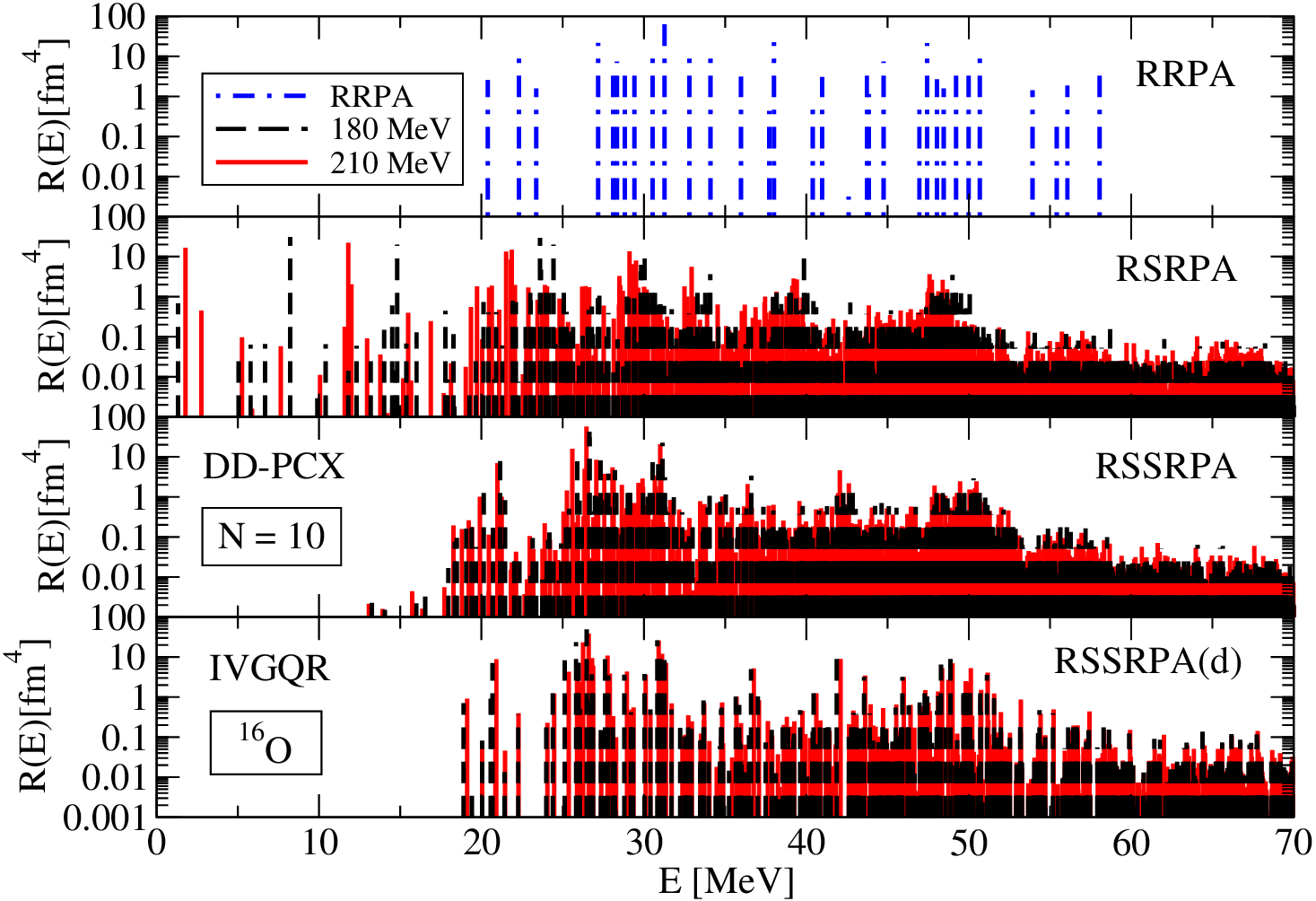}
\caption{Discrete IVGQR strength distribution calculated in the RSRPA framework for \textsuperscript{16}O.}
\label{fig:IVGQR:Discrete}
\end{figure}

In the final part of this analysis, we examine the IVGQR response for $^{16}$O using the relativistic SRPA developed in this work. Figures~\ref{fig:IVGQR} and \ref{fig:IVGQR:Discrete} display the corresponding strength distributions, obtained through Lorentzian folding and in discrete form, respectively. Overall, the features of the IVGQR spectrum resemble those observed for the ISGQR. However, the discrepancy between the RSRPA and its diagonal approximation is noticeably smaller in the isovector channel, suggesting that the interaction among \(2p\)--\(2h\) configurations has a less significant impact on the structure of the IVGQR. This conclusion is further supported by the subtracted calculations, i.e., both RSSRPA and RSSRPA(d) display weak sensitivity to variations in the energy cutoff, with the transition strength distribution exhibiting reasonable convergence already at excitation energies beyond $\approx$150 MeV.

\section{Conclusions}\label{sec:conclusions}

In this study, the relativistic second random-phase approximation and its subtracted extension are developed and applied to investigate collective excitations in finite nuclei. By adding \(2p\text{--}2h\) configurations to the standard \(1p\text{--}1h\) and antiparticle–hole spaces, this approach captures complex mixing and the fragmentation of strength that are inaccessible in conventional RRPA. The subtraction procedure in RSSRPA removes double counting, improves stability with respect to the configuration space, and preserves the low-energy limit of RPA.

The first calculations for $^{16}$O showed that the inclusion of \(2p\text{--}2h\) correlations systematically lowers ISGMR and ISGQR centroid energies toward the experimental range and enhances fragmentation of the overall transition strength.
Analysis of both isoscalar and isovector modes shows that the RSSRPA results in improved convergence and yields centroid energies, widths, and strength distributions in better agreement with experiment than the un-subtracted scheme.
Remaining excess in the monopole energy-weighted sum rule suggests that correlations beyond \(2p\text{--}2h\) and improvements to the ground-state description are still important, i.e., the degree of correlations already present in the reference state. As discussed in~\cite{PhysRevC.87.054330}, neglecting such correlations in the relativistic Hartree approximation can lead to an overestimation of excitation energies, EWSR and not fully reproduced fragmentation pattern. This highlights the importance of consistently correlated ground states in the theoretical modeling of nuclear collective excitations.

Within the current computational implementation, the R(S)SRPA framework is presently limited to closed-shell (magic) nuclei. 
Future developments will require an extension toward a quasiparticle-based formulation combined with partial eigenvalue solvers in order to address medium-mass and heavy nuclei in a computationally feasible manner. 
The present framework can also be further extended through improved energy density functionals and the inclusion of pairing correlations. 
The employed variational basis furthermore provides a route toward a fully microscopic description of nuclear response at finite temperature and in charge-exchange channels, as well as toward applications in modeling nuclear decays and processes of astrophysical relevance.

\begin{acknowledgments}
We thank D. Gambacurta, P. Papakonstantinou, J. Kvasil, and A. Repko for helpful discussions and advice that contributed to the development of the second random phase approximation models. This research was supported by the Croatian Science Foundation under the project Relativistic Nuclear Many-Body Theory in the Multimessenger Observation Era (HRZZ-IP-2022-10-7773). This work was supported by the project “Implementation of cutting-edge research and its application as part of the Scientific Center of Excellence for Quantum and Complex Systems, and Representations of Lie Algebras“, Grant No. PK.1.1.10.0004, co-financed by the European Union through the European Regional Development Fund - Competitiveness and Cohesion Programme 2021-2027. G.S. acknowledges the support provided by SERB
(DST), Govt. of India under SIR/2022/000566.  
\end{acknowledgments}

\bibliographystyle{apsrev4-2}
\bibliography{references}
\end{document}